\documentclass[a4paper,12pt]{article}
\usepackage{amsmath,amssymb,amsthm}

\newtheorem{lemma}{Lemma}
\newtheorem{theorem}{Theorem}

\theoremstyle{definition}

\theoremstyle{remark}

\newcommand{\beq}{\begin{eqnarray}}
\newcommand{\eeq}{\end{eqnarray}}
\newcommand{\beqnn}{\begin{eqnarray*}}
\newcommand{\eeqnn}{\end{eqnarray*}}
\newcommand{\rd}{\partial}

\newcommand{\ZZ}{\mathbf{Z}}
\newcommand{\bse}{\boldsymbol{e}}
\newcommand{\bss}{\boldsymbol{s}}
\newcommand{\bst}{\boldsymbol{t}}
\newcommand{\bszero}{\boldsymbol{0}}

\newcommand{\calS}{\mathcal{S}}

\begin{document}

\title{Universal Whitham hierarchy, \\
dispersionless Hirota equations and \\
multi-component KP hierarchy}

\author{Kanehisa Takasaki\\
Graduate School of Human and Environmental Studies,\\
Kyoto University\\
Yoshida, Sakyo, Kyoto, 606-8501, Japan\\
E-mail: takasaki@math.h.kyoto-u.ac.jp
\\\\
Takashi Takebe\\
Department of Mathematics, Ochanomizu University\\
Otsuka 2-1-1, Bunkyo-ku, Tokyo, 112-8610, Japan\\ 
E-mail: takebe@math.ocha.ac.jp}

\date{}
\maketitle

\begin{abstract}
The goal of this paper is to identify 
the universal Whitham hierarchy of genus zero 
with a dispersionless limit of the multi-component 
KP hierarchy.  To this end, the multi-component 
KP hierarchy is (re)formulated to depend on 
several discrete variables called ``charges''.  
These discrete variables play the role of 
lattice coordinates in underlying Toda field equations.  
A multi-component version of the so called 
differential Fay identity are derived from 
the Hirota equations of the $\tau$-function of 
this ``charged'' multi-component KP hierarchy.  
These multi-component differential Fay identities 
have a well-defined dispersionless limit 
(the dispersionless Hirota equations).  
The dispersionless Hirota equations turn out 
to be equivalent to the Hamilton-Jacobi equations 
for the $S$-functions of the universal Whitham 
hierarchy.  The differential Fay identities 
themselves are shown to be a generating functional 
expression of auxiliary linear equations for 
scalar-valued wave functions of the multi-component 
KP hierarchy.  
\end{abstract}
\bigskip

\begin{flushleft}
arXiv:nlin.SI/0608068
\end{flushleft}
\newpage

\section{Introduction}

Nowadays most soliton equations are known to be 
(re)formulated in the form of Hirota equations 
for $\tau$-functions \cite{JM-83}.  
If such a soliton equation has a dispersionless limit 
({\em dispersionless integrable system}), 
a natural question will be whether the Hirota equations 
have a reasonable dispersionless limit 
({\em dispersionless Hirota equations}).  
This issue has been studied for the following 
relatively small number of cases (because 
the dispersionless integrable systems themselves 
are, at least currently, rare species).  
\begin{itemize}
\item[1.] The first example of dispersionless Hirota 
equations was discovered for the dispersionless KP 
hierarchy \cite{TT-review}.   This dispersionless 
Hirota equation was obtained as a dispersionless limit 
of the so called {\em differential Fay identity} 
\cite{AvM-92} (which actually turned out to be equivalent 
to the KP hierarchy itself).  Another derivation 
using a Cauchy kernel was later developed \cite{CK-95}.  

\item[2.] Dispersionless Hirota equations are 
also known for the dispersionless Toda hierarchy, 
and applied to Laplacian growth 
\cite{WZ-00,MWWZ-00,KKMWWZ-01,Zabrodin-01}, 
WDVV associativity equations \cite{BMRWZ-01}, 
string field theory \cite{BR-SFT,BKR-SFT,BS-SFT,BSSST-SFT}, 
etc.  Moreover, a connection with some notions of 
complex analysis (the Grunsky coefficients and 
the Faber polynomials) was pointed out 
\cite{Takhtajan-01,Teo-03}.  Recently, 
a set of Fay-like identities for the Toda hierarchy 
were derived and shown to reduce to the dispersionless 
Hirota equations in the dispersionless limit \cite{Teo-06}.  

\item[3.] A few other exotic soliton equation 
and some problems in complex analysis have been 
studied in the context of dispersionless Hirota 
equations 
\cite{BK-dBKP,Takasaki-2BKP,CT-dBKP,TTZ-Loewner}.
\end{itemize}
The goal of this paper is to enlarge this list 
to include a multi-component generalization of 
the KP hierarchy ({\em multi-component KP hierarchy}) 
\cite{Sato-81,SS-83,DJKM-III,Dickey-book,KvdL-mcKP}.  

Unlike the aforementioned cases, the multi-component 
KP hierarchy is usually formulated in a matrix form 
with matrix (pseudo)differential operators and vector 
(or matrix) valued wave functions.  
This yields technical difficulties when one naively 
attempts to give a prescription of dispersionless limit 
in the Lax formalism.   As regards the multi-component 
KP hierarchy, thus, defining a reasonable 
dispersionless limit itself has been considered 
a tough problem.  

These difficulties can be circumvented in two ways.  
One way is to resort to a {\em scalar} Lax formalism 
based on scalar operators and wave functions 
\cite{Takasaki-SISSA05}. 
A prototype of such a scalar Lax formalism 
can be seen in the Toda hierarchy, which is 
equivalent to the two-component KP hierarchy 
that depends an extra discrete variable $s$ 
(i.e., the lattice coordinate)  \cite{UT-84}, 
and nevertheless has a scalar Lax formalism 
(in terms of difference operators in $s$).  
It is rather straightforward to generalize 
this point of view to the case with more than 
two components.  The other way is to seek for 
a prescription of dispersionless limit for 
Hirota equations rather than the Lax formalism.  
Actually, these two approaches lead to 
the same result.  In this paper, we primarily 
choose the route from Hirota equations, 
which eventually reaches the scalar Lax formalism 
as well.  

In the case of the $N+1$-component KP hierarchy, 
we can introduce $N$ discrete variables $s_1,\ldots,s_N$.  
These variables stand for charges of a state in 
the Fock space of an $N+1$-component charged 
free fermion system in the field theoretical formalism 
\cite{DJKM-III,KvdL-mcKP}.  In analogy with 
the aforementioned interpretation of 
the Toda hierarchy, this {\em charged} multi-component 
KP hierarchy may be thought of as a generalization 
of the Toda hierarchy as well.  

The dispersionless limit of this hierarchy 
turns out to be the universal Whitham hierarchy 
of genus zero.  The universal Whitham hierarchy 
was introduced, at all genera, by Krichever 
\cite{Krichever-94} as a universal framework 
for both dispersionless integrable systems and 
Whitham modulation equations.  The nonzero-genus cases 
were recently applied to Laplacian growth of 
multiply-connected domains \cite{KMWWZ-04,KMZ-05}.  
The zero-genus case, too, is the target of 
recent studies \cite{GMMA-03,MMMA-0509,MMMA-0510}, 
which have revealed interesting new aspects 
of this hierarchy.   One can see from our results 
that many of them stem from the multi-component 
KP hierarchy.  

This paper is organized as follows.   
Sections 2 is a review of the universal 
Whitham hierarchy of genus zero and some 
related equations.  Of particular importance 
is the Hamilton-Jacobi equations satisfied 
by the $S$-functions.   These equations play 
the role of auxiliary linear equations 
in usual soliton equations.  In Section 3, 
we show that these equations can be converted 
to a generating functional form, which we shall 
later identify with dispersionless Hirota equations.  
Section 4 presents a brief account of the $\tau$-function 
of the ``charged'' multi-component KP hierarchy 
and the Hirota equations in a generating functional 
bilinear form.  In Section 5, we derive a set of 
differential Fay identities for this hierarchy, 
and show that their dispersionless limit (namely, 
the dispersionless Hirota equations in this case) 
coincide with the equations obtained in Section 3.  
In Section 6, we show that these differential 
Fay identities are in fact a generating functional 
expression of auxiliary linear equations 
in the aforementioned scalar Lax formalism.

\section{Universal Whitham hierarchy of genus zero}

\subsection{Dynamical variables}

We consider the universal Whitham hierarchy of 
genus zero with $N+1$ marked points (or ``punctures'') 
on a Riemann sphere with coordinate $p$.  
One of these marked points is fixed at $p = \infty$, 
and the others are located at $p = q_1,\ldots,q_N$, 
which are part of dynamical variables.  
The other dynamical variables are the coefficients 
of Laurent series $z_0(p),z_1(p),\ldots,z_N(p)$ 
of the form 
\beq
\begin{aligned}
  z_0(p) &= p + \sum_{j=2}^\infty u_{0j}p^{-j+1}, \\
  z_\alpha(p) &= \frac{r_\alpha}{p - q_\alpha} 
    + \sum_{j=1}^\infty u_{\alpha j}(p - q_\alpha)^{j-1} 
    \quad (\alpha = 1,\ldots,N) 
\end{aligned}
\label{z(p)-def} 
\eeq
defined at the marked points.  

The hierarchy has $N+1$ sets of time variables 
$t_{0n}$ ($n = 1,2,\ldots$) and $t_{\alpha n}$ 
($\alpha = 1,\ldots,N$, $n = 0,1,\ldots$) 
attached to the marked points at $p = \infty$ 
and $p = q_\alpha$ ($\alpha = 1,\ldots,N$), 
respectively.  The lowest ones $t_{01},t_{10},
\ldots,t_{N0}$ play a special role.  For convenience, 
we introduce the auxiliary variable 
\beqnn
  t_{00} = - \sum_{\alpha=1}^N t_{\alpha 0}, 
\eeqnn
which is, of course, not an independent variable. 
Let $\rd_{\alpha n}$ denote the derivatives 
\beqnn
  \rd_{\alpha n} = \rd/\rd t_{\alpha n}. 
\eeqnn

\subsection{Lax and Zakharov-Shabat equations}

Time evolutions of the hierarchy are generated 
by the dispersionless Lax equations 
\beq
  \rd_{\alpha n}z_\beta(p) 
  = \{\Omega_{\alpha n}(p),\, z_\beta(p)\} 
\eeq
with respect to the Poisson bracket 
\beq
  \{f,g\} 
  = \frac{\rd f}{\rd p}\frac{\rd g}{\rd t_{01}} 
  - \frac{\rd f}{\rd t_{01}}\frac{\rd g}{\rd p}. 
\eeq
The ``Hamiltonians'' $\Omega_{\alpha n}(p)$ 
are defined as 
\beq
\begin{aligned}
  &\Omega_{0n}(p) 
  = \bigl(z_0(p)^n\bigr)_{(0,\ge 0)}, \quad 
  \Omega_{\alpha n}(p)
  = \bigl(z_\alpha(p)^n\bigr)_{(\alpha,>0)} 
  \quad (n = 1,2,\ldots), \\
  &\Omega_{\alpha 0}(p) 
  = - \log(p - q_\alpha), 
\end{aligned}
\eeq
where $(\quad)_{(0,\ge 0)}$ denotes the projection 
to nonnegative powers of $p$, and $(\quad)_{(\alpha,>0)}$ 
the projection to positive powers of $(p - q_\alpha)^{-1}$.  
In other words, 
\beq
\begin{aligned}
  z_0(p)^n &= \Omega_{0n}(p) + O(p^{-1})
  \quad (p \to \infty), \\
  z_\alpha(p)^n &= \Omega_{\alpha n}(p) + O(1) 
  \quad (p \to q_\alpha) 
\end{aligned}
\eeq
for $n \ge 1$.  In particular, 
\beqnn
  \Omega_{01}(p) = p, \quad 
  \Omega_{\alpha 1}(p) = \frac{r_\alpha}{p - q_\alpha}. 
\eeqnn

The Hamiltonians $\Omega_{\alpha n}(p)$ satisfy 
the dispersionless Zakharov-Shabat equations 
\beq
  \rd_{\beta n}\Omega_{\alpha m}(p) 
  - \rd_{\alpha m}\Omega_{\beta n}(p) 
  + \{\Omega_{\alpha m}(p),\,\Omega_{\beta n}(p)\} 
  = 0. 
\eeq
They can be converted to the equation 
\beq
  \omega \wedge \omega = 0 
\eeq
of the closed 2-form 
\beqnn
  \omega 
  = \sum_{n=1}^\infty d\Omega_{0n}(p)\wedge dt_{0n} 
  + \sum_{\alpha=1}^N\sum_{n=0}^\infty 
      d\Omega_{\alpha n}(p)\wedge dt_{\alpha n}. 
\eeqnn
By Darboux's theorem, $z_\beta(p)$ 
has a ``conjugate'' variable $\zeta_\beta(p)$ 
with which $\omega$ can be written in 
the ``canonical form'' 
\beq
  \omega = dz_\beta(p) \wedge d\zeta_\beta(p). 
\label{omega=dzdzeta}
\eeq
Consequently, $\zeta_\beta(p)$ satisfies 
the Lax equations 
\beq
  \rd_{\alpha n}\zeta_\beta(p) 
  = \{\Omega_{\alpha n}(p),\, \zeta_\beta(p)\} 
\eeq
of the same form as $z_\beta(p)$, alongside 
the canonical Poisson commutation relation 
\beq
  \{z_\beta(p),\zeta_\beta(p)\} = 1. 
\eeq
Thus $\zeta_\beta(p)$ amounts to 
the Orlov-Schulman functions in 
the dispersionless KP hierarchy.

\subsection{$S$-functions}

We can rewrite (\ref{omega=dzdzeta}) as 
\beqnn
  d(\theta + \zeta_\beta(p)dz_\beta(p)) = 0 
\eeqnn
where 
\beqnn
  \theta = \sum_{n=1}^\infty \Omega_{0n}(p)dt_{0n} 
    + \sum_{\alpha=1}^N\sum_{n=0}^\infty 
      \Omega_{\alpha n}dt_{\alpha n}. 
\eeqnn
This implies the existence of a function 
$\calS_\beta(p)$ such that 
\beqnn
  d\calS_\beta(p) = \theta + \zeta_\beta(p)dz_\beta(p) 
\eeqnn
or, more explicitly, 
\beq
  d\calS_\beta(p) 
  = \zeta_\beta(p)dz_\beta(p) 
  + \sum_{n=1}^\infty \Omega_{0n}(p)dt_{0n} 
    + \sum_{\alpha=1}^N\sum_{n=0}^\infty 
        \Omega_{\alpha n}(p)dt_{\alpha n}. 
\label{dScal-eq}
\eeq
$\calS_0(p)$ and $\calS_\beta(p)$ 
turn out to have Laurent expansion 
of the following form: 
\beq
\begin{aligned}
  \calS_0(p) &= \sum_{n=1}^\infty t_{0n}z_0(p)^n 
    + t_{00}\log z_0(p) 
    - \sum_{n=1}^\infty \frac{z_0(p)^{-n}}{n}v_{0n}, \\
  \calS_\beta(p) &= \sum_{n=1}^\infty t_{\beta n}z_\beta(p)^n 
    + t_{\beta 0}\log z_\beta(p) + \phi_\beta 
    - \sum_{n=1}^\infty \frac{z_\beta(p)^{-n}}{n}v_{\beta n}, 
\end{aligned}
\label{calS}
\eeq
where $v_{0n},v_{\alpha n}$ and $\phi_\beta$ are 
functions of the time variables.  

These $S$-functions can also be written as 
\beqnn
  \calS_0(p) = S_0(z_0(p)), \quad 
  \calS_\beta(p) = S_\beta(z_\beta(p)) 
\eeqnn
where $S_0(z)$ and $S_\beta(z)$ are defined as 
\beq
\begin{aligned}
  S_0(z) &= \sum_{n=1}^\infty t_{0n}z^n 
    + t_{00}\log z 
    - \sum_{n=1}^\infty \frac{z^{-n}}{n}v_{0n}, \\
  S_\beta(z) &= \sum_{n=1}^\infty t_{\beta n}z^n 
    + t_{\beta 0}\log z + \phi_\beta 
    - \sum_{n=1}^\infty \frac{z^{-n}}{n}v_{\beta n}. 
\end{aligned}
\label{S(z)-def}
\eeq
The latter functions, too, are called $S$-functions, 
which play a more fundamental role in the subsequent 
consideration.  

As (\ref{dScal-eq}) implies, differentiating 
(\ref{calS}) by $z_\beta(p)$ while 
leaving the time variables constant 
yields $\zeta_0(p)$ and $\zeta_\beta(p)$, 
which thus turns out to be Laurent series 
of the form 
\beq
\begin{aligned}
  \zeta_0(p) 
  &= \sum_{n=1}^\infty nt_{0n}z_0(p)^{n-1} 
    + \frac{t_{00}}{z_0(p)}
    + \sum_{n=1}^\infty z_0(p)^{-n-1}v_{0n}, \\
  \zeta_\beta(p) 
  &= \sum_{n=1}^\infty nt_{\alpha n}z_\beta(p)^{n-1}  
    + \frac{t_{\beta 0}}{z_\beta(p)} 
    + \sum_{n=1}^\infty z_\beta(p)^{-n}v_{\beta n}. 
\end{aligned}
\eeq
Similarly, differentiating (\ref{calS}) 
by the time variables gives rise to 
the following expressions of $\Omega_{\alpha n}(p)$'s: 
\beq
  \Omega_{0n}(p) &=& 
  \begin{cases}
  \displaystyle
  z_0(p)^n - \sum_{m=1}^\infty \frac{z_0(p)^{-m}}{m}\rd_{0n}v_{0m},\\
  \displaystyle 
  \rd_{0n}\phi_\beta 
  - \sum_{m=1}^\infty \frac{z_\beta(p)^{-m}}{m}\rd_{0n}v_{\beta m}, 
  \end{cases}
  \\
  \Omega_{\alpha n}(p) &=& 
  \begin{cases}
  \displaystyle 
  - \delta_{n0}\log z_0(p) 
  - \sum_{m=1}^\infty \frac{z_0(p)^{-m}}{m}\rd_{\alpha 0}v_{0m},\\
  \displaystyle 
  \delta_{\alpha\beta}\log z_\beta(p) + \rd_{\alpha 0}\phi_\beta 
  - \sum_{m=1}^\infty \frac{z_\beta(p)^{-m}}{m}\rd_{\alpha 0}v_{\beta m}. 
  \end{cases}
\eeq

\subsection{Hamilton-Jacobi equations}

Since $\Omega_{01}(p) = p$, the foregoing 
expression of $\Omega_{01}(p)$ implies 
that $z_0(p)$ and $z_\beta(p)$ satisfy 
the equations 
\beq
\begin{aligned}
  p &= z_0(p) 
      - \sum_{m=1}^\infty 
        \frac{z_0(p)^{-m}}{m}\rd_{01}v_{0m}, \\
  p &= \rd_{01}\phi_\beta 
      - \sum_{m=1}^\infty 
        \frac{z_\beta(p)^{-m}}{m}\rd_{01}v_{\beta m}. 
\end{aligned}
\eeq
In other words, the inverse functions 
$p = p_0(z)$ and $p = p_\beta(z)$ of 
$z = z_0(p)$ and $z = z_\beta(p)$ 
are given explicitly by 
\beq
\begin{aligned}
  p_0(z) 
  = z - \sum_{m=1}^\infty \frac{z^{-m}}{m}\rd_{01}v_{0m}
  = \rd_{01}S_0(z), \\
  p_\beta(z) 
  = \rd_{01}\phi_\beta 
    - \sum_{m=1}^\infty \frac{z^{-m}}{m}\rd_{01}v_{\beta m} 
  = \rd_{01}S_\beta(z). 
\end{aligned}
\label{p(z)-def}
\eeq
As a byproduct of these relations, 
we find an expression of $p_\beta$ and 
$r_\beta$ in terms of the coefficients 
of $S_\beta(z)$:  Since $z_\beta(p)$ is assumed 
to have Laurent expansion of the form 
\beqnn
  z_\beta(p) = \frac{r_\beta}{p - q_\beta} + O(1) 
\eeqnn
at $p = q_\beta$, the inverse function should 
have Laurent expansion of the form 
\beqnn
  p_\beta(z) = q_\beta + r_\beta z^{-1} + O(z^{-2}). 
\eeqnn
Comparing this with (\ref{z(p)-def}), 
we readily find that 
\beq
  q_\beta = \rd_{01}\phi_\beta, \quad 
  r_\beta = - \rd_{01}v_{\beta 1}. 
\label{qr=der-phi}
\eeq

We can now derive a system of Hamilton-Jacobi type 
for the $S$-functions.  This is achieved by 
substituting $p = p_\beta(z)$ ($\beta = 0,1,\ldots,N$) 
in (\ref{dScal-eq}).  The outcome of substitution 
is the equation 
\beq
  dS_\beta(z) 
  = S_\beta'(z)dz 
    + \sum_{n=1}^\infty \Omega_{0n}(p_\beta(z))dt_{0n} 
    + \sum_{\alpha=1}^N\sum_{n=0}^\infty 
      \Omega_{\alpha n}(p_\beta(z))dt_{\alpha n}, 
\eeq
which implies that 
\beqnn
  \rd_{\alpha n}S_\beta(z) = \Omega_{\alpha n}(p_\beta(z)). 
\eeqnn
Since $p_\beta(z)$ is related to the $S$-functions 
as (\ref{p(z)-def}) shows, we eventually obtain 
the Hamilton-Jacobi equations
\beq
  \rd_{\alpha n}S_\beta(z) 
  = \Omega_{\alpha n}(\rd_{01}S_\beta(z)). 
\label{HJ-eq}
\eeq

The lowest Hamilton-Jacobi equations 
have a special meaning.  Whereas 
the equation for $t_{01}$ is trivial, 
those for $t_{\alpha 0}$ ($\alpha =1,\ldots,N$) 
read 
\beq
  \rd_{\alpha 0}S_\beta(z) 
  = - \log(\rd_{01}S_\beta(z) - q_\alpha), 
\label{HJ-eq-n=0}
\eeq
which can be solved for $p_\beta(z) 
= \rd_{01}S_\beta(z)$ as 
\beq
  p_\beta(z) 
  = e^{-\rd_{\alpha 0}S_\beta(z)} + q_\alpha. 
\label{HJ-eq-n=0bis}
\eeq
In particular, choosing $\beta = \alpha$ 
and extracting the coefficient of $z^{-1}$, 
we find another expression of $r_\alpha$: 
\beq
  r_\alpha = e^{-\rd_{\alpha 0}\phi_\alpha}. 
\label{r=exp-phi}
\eeq

These Hamilton-Jacobi equations 
characterize the $S$-functions 
in their own terms.  One can recover 
the previous setting of the universal 
Whitham hierarchy by just changing 
variables from $z$ to $p$.  In this sense, 
the two systems are two different pictures 
of the same system, and one can move from 
one picture to the other by (\ref{z(p)-def}) 
and (\ref{p(z)-def}).

\subsection{$F$-function}

Following Mart\'{\i}nez Alonso, Ma\~{n}as 
and Medina \cite{MMMA-0510}, 
we define the $F$-function (the logarithm 
of the dispersionless $\tau$-function 
\cite{Krichever-94}) to be a solution of 
the following equations: 
\beq
\begin{aligned}
&\rd_{0n}F = v_{0n}, \quad 
  \rd_{\alpha n}F = v_{\alpha n} 
  \quad (n = 1,2,\ldots), \\
&\rd_{\alpha 0}F 
    = - \phi_\alpha 
      + \sum_{\beta=1}^\alpha t_{\beta 0}\log(-1) 
  \quad (\alpha = 1,\ldots,N), 
\end{aligned}
\label{F-def}
\eeq
where $\log(-1)$ is understood to be equal 
to, say, $\pi i$, though the choice of 
the branch is irrelevant in the final result. 
Now we can rewrite (\ref{S(z)-def}) as 
\beq
\begin{aligned}
  S_0(z) 
  &= \sum_{n=1}^\infty t_{0n}z^n + t_{00}\log z 
   -  D_0(z)F, \\
  S_\alpha(z) 
  &= \sum_{n=1}^\infty t_{\alpha n}z^n 
     + t_{\alpha 0}\log z + \phi_\alpha 
     - D_\alpha(z)F, 
\end{aligned}
\label{S(z)-F}
\eeq
where $D_0(z)$ and $D_\alpha(z)$ denotes 
the following differential operators:  
\beqnn
  D_0(z) 
  = \sum_{n=1}^\infty \frac{z^{-n}}{n}\rd_{0n}, \quad 
  D_\alpha(z) 
  = \sum_{n=1}^\infty \frac{z^{-n}}{n}\rd_{\alpha n}. 
\eeqnn

It should be mentioned that the last part of
(\ref{F-def}) is slightly different from that of 
Ma\~{n}as et al. \cite{MMMA-0510}.  Their proof 
of consistency of the defining equations, however, 
persists to be valid in this form as well.  
We have modified their definition so as to 
match the fermionic formula of the tau functions 
by Date et al. \cite{DJKM-III}.

\section{Dispersionless Hirota equations 
as generating functional form of 
Hamilton-Jacobi equations}

\subsection{Faber polynomials}

Following Teo's idea \cite{Teo-03} 
developed for the case of the dispersionless 
KP and Toda hierarchies, we now consider 
the notion of Faber polynomials of 
$p_0(z)$ and $p_\alpha(z)$, and 
show that they actually coincide 
with $\Omega_{0n}(p)$ and $\Omega_{\alpha n}(p)$.  

The Faber polynomials $\Phi_{0n}(p)$ and 
$\Phi_{\alpha n}(p)$ of $p_0(z)$ and 
$p_\alpha(z)$ are defined by the following 
generating functions:
\beq
\begin{aligned}
  \log\frac{p_0(z) - q}{z} 
  &= - \sum_{n=1}^\infty 
       \frac{z^{-n}}{n}\Phi_{0n}(q), \\
  \log\frac{q - p_\alpha(z)}{q - q_\alpha} 
  &= - \sum_{n=1}^\infty 
       \frac{z^{-n}}{n}\Phi_{\alpha n}(q). 
\end{aligned}
\eeq
Rewriting the left hand side as 
\beqnn
  \log\frac{p_0(z) - q}{z} 
  = \log\frac{p_0(z)}{z} 
    + \log\Bigl(1 - \frac{q}{p_0(z)}\Bigr) 
\eeqnn
and 
\beqnn
  \log\frac{q - p_\alpha(z)}{q - q_\alpha} 
  = \log\Bigl(1 - \frac{p_\alpha(z) - q_\alpha}{q - q_\alpha}\Bigr)
\eeqnn
and recalling that 
\beqnn
  p_0(z) = z + O(z^{-1}), \quad 
  p_\alpha(z) = q_\alpha + O(z^{-1}), 
\eeqnn
we readily see that 
\begin{itemize}
\item[1)] $\Phi_{0n}(q)$ is a monic polynomial of $q$, and 
\item[2)] $\Phi_{\alpha n}(q)$ is a polynomial of 
$(q - q_\alpha)^{-1}$ with no constant term. 
\end{itemize}

\begin{lemma}
The Faber polynomials coincide with 
$\Omega_{0n}(p)$ and $\Omega_{\alpha n}(p)$: 
\beq
  \Phi_{0n}(q) = \Omega_{0n}(q), \quad 
  \Phi_{\alpha n}(q) = \Omega_{\alpha n}(q). 
\eeq
\end{lemma}

\proof 
A more explicit expression of these polynomials 
can be obtained by differentiating the generating 
functions by $z$ as 
\beqnn
\begin{aligned}
  \frac{p_0'(z)}{p_0(z) - q} - \frac{1}{z} 
  &= \sum_{n=1}^\infty z^{-n-1}\Phi_{0n}(q), \\
  \frac{p_\alpha'(z)}{p_\alpha(z) - q} 
  &= \sum_{n=1}^\infty z^{-n-1}\Phi_{\alpha n}(q) 
\end{aligned}
\eeqnn
and extracting the Laurent coefficients by 
contour integrals.  As regards $\Phi_{0n}(z)$, 
this yields the contour integral formula 
\beqnn
  \Phi_{0n}(q) 
  = \frac{1}{2\pi i}\oint\frac{z^np_0'(z)dz}{p_0(z) - q} 
  = \frac{1}{2\pi i}\oint_{|p|=R}\frac{z_0(p)^ndp}{p - q}, 
\eeqnn
where the contour $|p| = R$ in the second integral 
is understood to be sufficiently large and 
to encircle $q$ anti-clockwise; the contour in 
the first integral is its image under 
the mapping $p \mapsto z_0(p)$.  
The last contour integral is nothing but 
the polynomial part of $z_0(p)^n$ 
evaluated at $q$, i.e., $\Omega_{0n}(q)$.  
In much the same way, we have the contour integral 
formula 
\beqnn
  \Phi_{\alpha n}(q) 
  = \frac{1}{2\pi i}\oint\frac{z^np_\alpha'(z)dz}{p_\alpha(z) - q} 
  = \frac{1}{2\pi i}\oint_{|p-q_\alpha|=r}
    \frac{z_\alpha(p)^ndp}{p - q}, 
\eeqnn
where the contour $|p-q_\alpha| = r$ in the second integral 
is chosen to be sufficiently small and to encircle 
$q_\alpha$ clockwise, leaving $q$ outside.  
This contour integral leaves negative powers 
of the Laurent expansion of $z_\alpha(p)^n$ 
at $p = q_\alpha$, hence coincides with 
$\Omega_{\alpha n}(q)$.  
\qed

Bearing this interpretation of $\Omega_{\alpha n}(p)$'s 
in mind, we turn to the Hamilton-Jacobi equations 
(\ref{HJ-eq}).   The goal is to transform 
these equations, more precisely, those for 
$n \ge 1$, into a generating functional form.

\subsection{First subset of Hamilton-Jacobi equations}

Let us first consider the equations 
for $\alpha = 0$, $\beta = 0$ and $n \ge 1$: 
\beq
  \rd_{0n}S_0(z) 
  = \Omega_{0n}(\rd_{01}S_0(z)) 
  = \Omega_{0n}(p_0(z)). 
\label{HJ-eq1}
\eeq

\begin{lemma}
(\ref{HJ-eq1}) is equivalent to 
\beq
  \log\frac{p_0(z) - p_0(w)}{z - w} = D_0(z)D_0(w)F 
\label{HJ-geneq1}
\eeq
\end{lemma}

\proof
Substituting $q = p_0(w)$ in the generating function 
of the Faber polynomials $\Phi_{0n}(q) = \Omega_{0n}(q)$ 
yields the identity 
\beqnn
  \log\frac{p_0(z) - p_0(w)}{z} 
  = - \sum_{n=1}^\infty 
      \frac{z^{-n}}{n}\Omega_{0n}(p_0(w)). 
\eeqnn
Therefore (\ref{HJ-eq1}) can be cast into 
the generating functional form 
\beqnn
  \log\frac{p_0(z) - p_0(w)}{z}
  = - \sum_{n=1}^\infty 
      \frac{z^{-m}}{m}\rd_{0n}S_0(w). 
\eeqnn
(\ref{S(z)-F}) implies that $\rd_{0n}S_0(w)$ 
on the right hand side can be expressed as 
\beqnn
  \rd_{0n}S_0(w) 
  = w^n - \sum_{m=1}^\infty 
          \frac{w^{-m}}{m}\rd_{0n}\rd_{0m}F. 
\eeqnn
Thus the generating functional equation 
can be rewritten as 
\beqnn
\begin{aligned}
  \log\frac{p_0(z) - p_0(w)}{z}
  &= - \sum_{n=1}^\infty \frac{z^{-n}}{n} 
       \Bigl(w^n - \sum_{m=1}^\infty 
         \frac{w^{-m}}{m}\rd_{0n}\rd_{0m}F \Bigr) \\
  &= \log\Bigl(1 - \frac{w}{z}\Bigr) + D_0(z)D_0(w)F. 
\end{aligned}
\eeqnn
Moving the first term on the right hand side 
to the left hand side, we obtain (\ref{HJ-geneq1}) 
as a generating functional form of (\ref{HJ-eq1}).  
\qed

\subsection{Second subset of Hamilton-Jacobi equations}

The second subset of equations are those for 
$\alpha = 1,\ldots,N$, $\beta = 0$, and $n \ge 1$: 
\beq
\begin{aligned}
  \rd_{\alpha n}S_0(z) 
  = \Omega_{\alpha n}(\rd_{01}S_0(z)) 
  = \Omega_{\alpha n}(p_0(z)). 
\end{aligned}
\label{HJ-eq2}
\eeq

\begin{lemma}
(\ref{HJ-eq2}) is equivalent to 
\beq
  \log\frac{p_0(z) - p_\beta(w)}{p_0(z) - q_\beta} 
  = D_0(z)D_\beta(w)F. 
\label{HJ-geneq2}
\eeq
\end{lemma}

\proof
Substituting $q = p_0(w)$ in the generating function 
of the Faber polynomials $\Phi_{\alpha n}(q) 
= \Omega_{\alpha n}(q)$ leads to the identity 
\beqnn
  \log\frac{p_0(w) - p_\alpha(z)}{p_0(w) - q_\alpha} 
  = - \sum_{n=1}^\infty \frac{z^{-n}}{n}\Omega_{\alpha n}(p_0(w)). 
\eeqnn
Therefore (\ref{HJ-eq2}) can be converted 
to the generating functional form 
\beqnn
  \log\frac{p_0(w) - p_\alpha(z)}{p_0(w) - q_\alpha} 
  = - \sum_{n=1}^\infty \rd_{\alpha n}S_0(w). 
\eeqnn
(\ref{S(z)-F}) implies that $\rd_{\alpha n}S_0(w)$ 
on the right hand side can be rewritten as 
\beqnn
  \rd_{\alpha n}S_0(w) 
  = - \sum_{m=1}^\infty \frac{w^{-m}}{m}\rd_{\alpha n}\rd_{0m}F. 
\eeqnn
Thus the generating functional equation 
turns into (\ref{HJ-eq2}).  
\qed

\subsection{Third subset of Hamilton-Jacobi equations}

Let us now consider the equations for 
$\alpha = 0$, $\beta = 1,\ldots,N$ and $n \ge 1$: 
\beq
  \rd_{0n}S_\beta(z) 
  = \Omega_{0n}(\rd_{01}S_\beta(z)) 
  = \Omega_{0n}(p_\beta(z)). 
\label{HJ-eq3}
\eeq
Unlike the previous two cases, 
we now have to use (\ref{HJ-eq-n=0}) 
to rewrite these Hamilton-Jacobi equations.  

\begin{lemma}
If (\ref{HJ-eq-n=0}) is satisfied, 
(\ref{HJ-eq3}) is equivalent to 
(\ref{HJ-geneq2}).  
\end{lemma}

\proof
As in the case of (\ref{HJ-eq1}), we can convert 
(\ref{HJ-eq3}) to the generating functional form 
\beqnn
  \log\frac{p_0(z) - p_\beta(w)}{z} 
  = - \sum_{n=1}^\infty 
      \frac{z^{-n}}{n}\rd_{0n}S_\beta(w). 
\eeqnn
By (\ref{S(z)-F}), we can rewrite $\rd_{0n}S_\beta(w)$ 
on the right hand side as 
\beqnn
  \rd_{0n}S_\beta(w) 
  = - \rd_{0n}\rd_{\beta 0}F 
    + \sum_{m=1}^\infty 
      \frac{w^{-m}}{m}\rd_{0n}\rd_{\beta m}F, 
\eeqnn
This leads to an equation of the form 
\beqnn
  \log\frac{p_0(z) - p_\beta(w)}{z} 
  = \sum_{n=1}^\infty \frac{z^{-n}}{n}\rd_{0n}\rd_{\beta 0}F 
  + \sum_{n,m=1}^\infty \frac{z^{-n}w^{-m}}{nm} 
    \rd_{0n}\rd_{\beta m}F. 
\eeqnn
The first sum on the right hand side 
can be evaluated as 
\beqnn
\begin{aligned}
  \sum_{n=1}^\infty \frac{z^{-n}}{n}\rd_{0n}\rd_{\beta 0}F 
  &= \rd_{\beta 0}\sum_{n=1}^\infty 
       \frac{z^{-n}}{n}\rd_{0n}F \\
  &= \rd_{\beta 0}(- S_0(z) + t_{00}\log z) \\
  &= - \rd_{\beta 0}S_0(z) - \log z. 
\end{aligned}
\eeqnn
If (\ref{HJ-eq-n=0}) is satisfied, 
we can rewrite $\rd_{\beta_0}S_0(z)$ as 
\beqnn
  \rd_{\beta 0}S_0(z) = - \log(p_\beta(z) - q_\beta) 
\eeqnn
and obtain (\ref{HJ-geneq2}). 
\qed

\subsection{Fourth subset of Hamilton-Jacobi equations} 

The fourth subset consists of the Hamilton-Jacobi 
equations for $\alpha = \beta = 1,\ldots,N$ 
and $n \ge 1$: 
\beq
\begin{aligned}
  \rd_{\alpha n}S_\alpha(z) 
  = \Omega_{\alpha n}(\rd_{01}S_\alpha(z)) 
  = \Omega_{\alpha n}(p_\alpha(z)). 
\end{aligned}
\label{HJ-eq4}
\eeq

\begin{lemma}
If (\ref{HJ-eq-n=0}) are satisfied, 
(\ref{HJ-eq4}) is equivalent to 
\beq
  \log\frac{p_\alpha(w) - p_\alpha(z)}
  {(z - w)(p_\alpha(z) - q_\alpha)(p_\alpha(w) - q_\alpha)}
  = \rd_{\alpha 0}\phi_\alpha + D_\alpha(z)D_\alpha(w)F 
\label{HJ-geneq3}
\eeq
\end{lemma}

\proof
As in the case of (\ref{HJ-eq2}), 
we convert (\ref{HJ-eq4})  to 
the generating functional form 
\beqnn
  \log\frac{p_\alpha(w) - p_\alpha(z)}{p_\alpha(w) - q_\alpha}
 = - \sum_{n=1}^\infty \frac{z^{-n}}{n}\rd_{\alpha n}S_\alpha(w) 
\eeqnn
and use (\ref{S(z)-F}) to rewrite $\rd_{\alpha n}S_\alpha(w)$ 
on the right hand side as 
\beqnn
  \rd_{\alpha n}S_\alpha(w) 
  = w^n - \rd_{\alpha n}\rd_{\alpha 0}F  
    - \sum_{m=1}^\infty 
      \frac{w^{-m}}{m}\rd_{\alpha n}\rd_{\alpha m}F. 
\eeqnn
This leads us to an equation of the form 
\beqnn
\begin{aligned}
  &\log\frac{p_\alpha(w) - p_\alpha(z)}{p_\alpha(w) - q_\alpha} \\
  &= \log\Bigl(1 - \frac{w}{z}\Bigr) 
    + \sum_{n=1}^\infty \frac{z^{-n}}{n}\rd_{\alpha n}\rd_{\alpha 0}F 
    + \sum_{n,m=1}^\infty 
      \frac{z^{-n}w^{-m}}{nm} \rd_{\alpha n}\rd_{\alpha 0}F. 
\end{aligned}
\eeqnn
The first sum on the right hand side can be 
evaluated as 
\beqnn
\begin{aligned}
  \sum_{n=1}^\infty \frac{z^{-n}}{n}\rd_{\alpha n}\rd_{\alpha 0}F 
  &=  \rd_{\alpha 0}\Bigl(\sum_{n=1}^\infty 
       \frac{z^{-n}}{n}\rd_{\alpha n}F\Bigr) \\
  &=  \rd_{\alpha 0}\Bigl(-S_\alpha(z) 
       + \sum_{n=1}^\infty t_{\alpha n}z^n 
       + t_{\alpha 0}\log z + \rd_{\alpha 0}\phi_\alpha\Bigr) \\
  &= - \rd_{\alpha 0}S_\alpha(z) + \log z 
     + \rd_{\alpha 0}\phi_\alpha. 
\end{aligned}
\eeqnn
As regards $\rd_{\alpha 0}S_\alpha(z)$, 
(\ref{HJ-eq-n=0}) implies that 
\beqnn
  \rd_{\alpha 0}S_\alpha(z) = - \log(p_\alpha(z) - q_\alpha). 
\eeqnn
We now move all logarithmic terms to the left hand side, 
and obtain (\ref{HJ-geneq3}) as a generating functional 
form of (\ref{HJ-eq4}).  
\qed

\subsection{Fifth subset of Hamilton-Jacobi equations}

The fifth subset of equations are those 
for $\alpha,\beta = 1,\ldots,N$, 
$\alpha \not= \beta$, and $n \ge 1$: 
\beq
  \rd_{\alpha n}S_\beta(z) 
  = \Omega_{\alpha n}(\rd_{01}S_\beta(z)) 
  = \Omega_{\alpha n}(p_\beta(z)) 
  \quad (\alpha \not= \beta). 
\label{HJ-eq5}
\eeq

\begin{lemma}
If (\ref{HJ-eq-n=0}) is satisfied, 
(\ref{HJ-eq5}) is equivalent to 
\beq
  \log\frac{p_\beta(w) - p_\alpha(z)}
  {(p_\beta(w) - q_\alpha)(p_\alpha(z) - q_\beta)} 
  = \rd_{\beta 0}\phi_\alpha 
    + D_\alpha(z)D_\beta(w)F 
 \label{HJ-geneq4}
\eeq
\end{lemma}

We first convert these equations to 
the generating functional form 
\beqnn
  \log\frac{p_\beta(w) - p_\alpha(z)}{p_\beta(w) - q_\alpha} 
  = - \sum_{n=1}^\infty \frac{z^{-n}}{n}\rd_{\alpha n}S_\beta(w)
\eeqnn
and use (\ref{S(z)-F}) to rewrite the derivatives 
on the right hand side as 
\beqnn
  \rd_{\alpha n}S_\beta(w) 
  = - \rd_{\alpha n}\rd_{\beta 0}F 
    - \sum_{m=1}^\infty 
      \frac{w^{-m}}{m}\rd_{\alpha n}\rd_{\beta m}F. 
\eeqnn
The outcome is an equation of the form 
\beqnn
  \log\frac{p_\beta(w) - p_\alpha(z)}{p_\beta(w) - q_\alpha} 
  = \sum_{n=1}^\infty \frac{z^{-n}}{n}\rd_{\alpha n}\rd_{\beta 0}F 
    + \sum_{n,m=1}^\infty \frac{z^{-n}w^{-m}}{nm} 
      \rd_{\alpha n}\rd_{\beta m}F.  
\eeqnn
The first sum on the right hand side can be 
evaluated in the same way as the previous case, 
namely, 
\beqnn
\begin{aligned}
  \sum_{n=1}^\infty \frac{z^{-n}}{n}\rd_{\alpha n}\rd_{\beta 0}F 
  &= \rd_{\beta 0}\Bigl(\sum_{n=1}^\infty 
       \frac{z^{-n}}{n}\rd_{\alpha n}F\Bigr)\\
  &= -\rd_{\beta 0}S_\alpha(z) + \rd_{\beta 0}\phi_0 \\
  & = \log(p_\alpha(z) - q_\beta) + \rd_{\beta 0}\phi_\alpha, 
\end{aligned}
\eeqnn
where (\ref{HJ-eq-n=0}) has been used in such a form as 
\beqnn
  \rd_{\beta 0}S_\alpha(z) = - \log(p_\alpha(z) - q_\beta). 
\eeqnn
Consequently, we have (\ref{HJ-geneq4}) as 
a generating functional form of (\ref{HJ-eq5}).  
\qed

\subsection{Dispersionless Hirota equations}

We have thus converted the $n \not= 0$ sector 
of the full Hamilton-Jacobi equations (\ref{HJ-eq}) 
to the four generating functional equations  
(\ref{HJ-geneq1}), (\ref{HJ-geneq2}), 
(\ref{HJ-geneq3}) and (\ref{HJ-geneq4}).  
Since the $n = 0$ sector (\ref{HJ-eq-n=0}) of 
the Hamilton-Jacobi equations is not included 
therein, these generating functional equations 
are still incomplete. 

As the following theorem shows, we can modify 
these generating functional equations 
to include (\ref{HJ-eq-n=0}) as well. 
The {\em complete} generating functional equations, 
(\ref{dHirota-eq1})--(\ref{dHirota-eq4}), 
resemble the dispersionless Hirota equations 
for the KP and Toda hierarchies.  As we shall 
show later, these equations can be derived from 
bilinear equations of the $\tau$-function of 
the multi-component KP hierarchy.  
For these reasons, we call these equations 
``dispersionless Hitota equations.''

\begin{theorem}
The full Hamilton-Jacobi equations (\ref{HJ-eq}) 
are equivalent to the following 
dispersionless Hirota equations: 
\beq
e^{D_0(z)D_0(w)F}
  &=& 1 - \frac{\rd_{01}(D_0(z) - D_0(w))F}{z-w}, 
\label{dHirota-eq1}\\
ze^{D_0(z)(\rd_{\alpha 0}+D_\alpha(w))F}
  &=& z - \rd_{01}(D_0(z) - \rd_{\alpha 0} - D_\alpha(w))F, 
\label{dHirota-eq2}\\
e^{(\rd_{\alpha 0}+D_\alpha(z))(\rd_{\alpha 0}+D_\alpha(w))F} 
  &=& - \frac{zw\rd_{01}(D_\alpha(z) - D_\alpha(w))F}{z-w}, 
\label{dHirota-eq3}\\
\epsilon_{\alpha\beta}
e^{(\rd_{\alpha 0}+D_\alpha(z))(\rd_{\beta 0}+D_\beta(w))F} 
  &=& - \rd_{01}(\rd_{\alpha 0} + D_\alpha(z) 
               - \rd_{\beta 0} - D_\beta(w))F, 
\label{dHirota-eq4}
\eeq
where $\epsilon_{\alpha\beta}$ is a sign factor 
defined as 
\beqnn
  \epsilon_{\alpha\beta} 
  = \epsilon_{\alpha\beta}(\bszero) 
  = \begin{cases}
    +1 & (\alpha \le \beta),\\
    -1 & (\alpha > \beta). 
    \end{cases}
\eeqnn
\end{theorem}

\proof
Because of the previous lemmas, we have only 
to show that the generating functional form 
of the Hamilton-Jacobi equations, upon supplemented 
by the $n = 0$ sector (\ref{HJ-eq-n=0}), is 
equivalent to the dispersionless Hirota equations. 
As we show below, the four generating functional 
equations (\ref{HJ-geneq1}), (\ref{HJ-geneq2}), 
(\ref{HJ-geneq3}) and (\ref{HJ-geneq4}) respectively 
correspond to the four parts (\ref{dHirota-eq1}), 
(\ref{dHirota-eq2}), (\ref{dHirota-eq3}) and (\ref{dHirota-eq4}) 
of the dispersionless Hirota equations.  

\paragraph*{(\ref{HJ-geneq1}) $\Leftrightarrow$ (\ref{dHirota-eq1})}
To derive (\ref{dHirota-eq1}) from (\ref{HJ-geneq1}), 
we exponentiate the latter as 
\beqnn
  p_0(z) - p_0(w) 
  = 1 - \frac{e^{D_0(z)D_0(w)F}}{z-w} 
\eeqnn
and substitute 
\beqnn
  p_0(z) = z - \rd_{01}D_0(z)F, \quad 
  p_0(w) = z - \rd_{01}D_0(w)F 
\eeqnn
(see (\ref{p(z)-def}) and (\ref{S(z)-F})). 
This procedure is obviously reversible.  

\paragraph*{(\ref{HJ-geneq2}) $\Rightarrow$ (\ref{dHirota-eq2})} 
We exponentiate (\ref{HJ-geneq2}) as 
\beqnn
  p_0(z) - p_\beta(w) 
  = (p_0(z) - q_\beta)e^{D_0(z)D_\beta(w)F} 
\eeqnn
and examine the both hand sides.  
As regards the left hand side, 
we substitute 
\beqnn
  p_0(z) = z - \rd_{01}D_0(z)F, \quad
  p_\beta(w) = - \rd_{01}\rd_{\beta 0}F 
               - \rd_{01}D_\beta(w)F
\eeqnn
and find that 
\beqnn
  \text{LHS} 
  = z - \rd_{01}(D_0(z) - \rd_{\beta 0} - D_\beta(w))F. 
\eeqnn
As regards the right hand side, 
we use the relation 
\beqnn
  p_0(z) - q_\beta 
  = e^{-\rd_{\beta 0}S_0(z)}
  = ze^{\rd_{\beta 0}D_0(z)F}  
\eeqnn
implied by (\ref{HJ-eq-n=0bis}) and (\ref{S(z)-F}).  
Thus we obtain (\ref{dHirota-eq2}).  

\paragraph*{(\ref{dHirota-eq2}) $\Rightarrow$ (\ref{HJ-geneq2})} 
Taking the limit as $w \to \infty$ in 
(\ref{dHirota-eq2}), we obtain the relation 
\beqnn
  z - \rd_{01}(D_0(z) - \rd_{\beta 0})F 
  = ze^{\rd_{\beta 0}D_0(z)F} 
  = e^{-\rd_{\beta 0}S_0(z)}. 
\eeqnn
Since (\ref{qr=der-phi}) and (\ref{F-def}) 
imply that 
\beqnn
  \rd_{01}\rd_{\beta 0}F = - q_\beta, \quad 
  p_0(z) = z - \rd_{01}D_0(z)F, 
\eeqnn
we can rewrite this relation as 
\beqnn
  p_0(z) - q_\beta = e^{-\rd_{\beta 0}S_0(z)}, 
\eeqnn
which is equivalent to one of the equations 
of (\ref{HJ-eq-n=0}).  Having this equation, 
we can recover (\ref{HJ-geneq2}) from (\ref{dHirota-eq2}).  

\paragraph*{(\ref{HJ-geneq3}) $\Rightarrow$ (\ref{dHirota-eq3})} 
We rewrite (\ref{HJ-geneq3}) as 
\beqnn
  \frac{p_\alpha(w) - p_\alpha(z)}{z - w} 
  = (p_\alpha(z) - q_\alpha)(p_\alpha(w) - q_\alpha) 
    \exp\bigl(\rd_{\alpha 0}\phi_\alpha 
              + D_\alpha(z)D_\alpha(w)F\bigr) 
\eeqnn
and examine the both hand sides.  
The left hand side can be processed 
in the same way as in the preceding cases: 
\beqnn
  \text{LHS} 
  =  \frac{\rd_{01}(D_\alpha(z)-D_\alpha(w))F}{z-w}. 
\eeqnn
As regards the right hand side, 
we use the relation 
\beqnn
  p_\alpha(z) - q_\alpha 
  = e^{-\rd_{\alpha 0}S_\alpha(z)} 
  = z^{-1}e^{-\rd_{\alpha 0}\phi_\alpha + \rd_{\alpha 0}D_\alpha(z)F} 
\eeqnn
and its counterpart for $p_\alpha(w) - q_\alpha$
to rewrite it as 
\beqnn
  \text{RHS} 
  = z^{-1}w^{-1}
     \exp\bigl(-\rd_{\alpha 0}\phi 
      + \rd_{\alpha 0}D_\alpha(z)F 
      + \rd_{\alpha 0}D_\alpha(w)F 
      + D_\alpha(z)D_\alpha(w)F \bigr). 
\eeqnn
Since (\ref{F-def}) implies that 
\beqnn
  \rd_{\alpha 0}\phi_\alpha 
  = - \rd_{\alpha 0}^2F + \log(-1), 
\eeqnn
we can further rewrite the foregoing 
expression as 
\beqnn
\begin{aligned}
  \text{RHS} 
  &= - z^{-1}w^{-1}
       \exp\bigl(\rd_{\alpha 0}^2F  
        + \rd_{\alpha 0}D_\alpha(z)F 
        + \rd_{\alpha 0}D_\alpha(w)F 
        + D_\alpha(z)D_\alpha(w)F \bigr) \\
  &= - z^{-1}w^{-1}
       e^{(\rd_{\alpha 0}+D_\alpha(z))
          (\rd_{\alpha 0}+D_\alpha(w))F}. 
\end{aligned}
\eeqnn
This completes the derivation of (\ref{dHirota-eq3}).  

\paragraph*{(\ref{dHirota-eq3}) $\Rightarrow$ 
(\ref{HJ-geneq3})}  
Taking the limit as $w \to \infty$ 
in (\ref{dHirota-eq3}) leads to the equation 
\beqnn
  \rd_{01}D_\alpha(z)F 
  = e^{(\rd_{\alpha 0}+D_\alpha(z))\rd_{\alpha 0}F}. 
\eeqnn
Since 
\beqnn
  \rd_{01}\phi_\alpha = q_\alpha, \quad 
  \rd_{\alpha 0}^2F 
    = -\rd_{\alpha 0}\phi + \log(-1), 
\eeqnn
we can rewrite the last equation as 
\beqnn
  p_\alpha(z) - q_\alpha 
  = e^{-\rd_{\alpha 0}\phi_\alpha 
       - \rd_{\alpha 0}D_\alpha(z)F} 
  = e^{-\rd_{\alpha 0}S_\alpha(z)}. 
\eeqnn
This is equivalent to one of the equations 
of (\ref{HJ-eq-n=0}).  We can thereby recover 
(\ref{HJ-geneq3}).  

\paragraph*{(\ref{HJ-geneq4}) $\Rightarrow$ 
(\ref{dHirota-eq4})} 
We first exponentiate (\ref{HJ-geneq4}) as 
\beqnn
  p_\beta(w) - p_\alpha(z) 
  = (p_\beta(w) - q_\alpha)(p_\alpha(z) - q_\beta) 
    e^{\rd_{\beta 0}\phi_\alpha + D_\alpha(z)D_\beta(w)F}. 
\eeqnn
Repeating almost the same calculations 
as the previous cases, we can rewrite 
the left hand side as 
\beqnn
  \text{LHS} 
  = \rd_{01}(\rd_{\alpha 0} + D_\alpha(z) 
           - \rd_{\beta 0} - D_\beta(w))F. 
\eeqnn
As regards the right hand side, we use the relation 
\beqnn
  p_\alpha(z) - q_\beta 
  = e^{-\rd_{\beta 0}S_\alpha(z)} 
  = e^{-\rd_{\beta 0}\phi_\alpha + \rd_{\beta 0}D_\alpha(z)F} 
\eeqnn
and its counterpart for $p_\beta(w) - q_\alpha$ 
implied by (\ref{HJ-eq-n=0bis}) and (\ref{S(z)-F}) 
to find that 
\beqnn
  \text{RHS} 
  = \exp\bigl(-\rd_{\alpha 0}\phi_\beta 
        - \rd_{\alpha 0}\rd_{\beta 0}F 
        + (\rd_{\alpha 0}+D_\alpha(z))
          (\rd_{\beta 0}+D_\beta(w))F \bigr). 
\eeqnn
On the other hand, (\ref{F-def}) implies that 
\beqnn
  \rd_{\alpha 0}\phi_\beta + \rd_{\alpha 0}\rd_{\beta 0}F 
  = \begin{cases}
    \log(-1) & (\alpha \le \beta),\\
    0  & (\alpha > \beta), 
    \end{cases}
\eeqnn
so that 
\beqnn
  \exp(-\rd_{\beta 0}\phi_\alpha 
       - \rd_{\alpha 0}\rd_{\beta 0}F) 
  = - \epsilon_{\alpha\beta}.  
\eeqnn
We can thus derive (\ref{dHirota-eq4}) from  
(\ref{HJ-geneq4}).  

\paragraph*{(\ref{dHirota-eq4}) $\Rightarrow$ 
(\ref{HJ-geneq4})} 

This is mostly parallel to the previous cases.  
We take the limit as $w \to \infty$ in 
(\ref{dHirota-eq4}).  The outcome is the equation 
\beqnn
  - \rd_{01}(\rd_{\alpha 0} + D_\alpha(z) 
           - \rd_{\beta 0})F 
  = \epsilon_{\alpha\beta}
    e^{(\rd_{\alpha 0}+D_\alpha(z))\rd_{\beta 0}F}. 
\eeqnn
It is now easy to see that the left hand side 
is equal to $p_\alpha(z) - q_\beta$.  
As regards the right hand side, 
we can use the aforementioned relation  
\beqnn
  \exp(-\rd_{\beta 0}\phi_\alpha 
       - \rd_{\alpha 0}\rd_{\beta 0}F) 
  = - \epsilon_{\alpha\beta}   
\eeqnn
once again to rewrite it as 
\beqnn
  \epsilon_{\alpha\beta}
  e^{\rd_{\beta 0}\rd_{\alpha 0}F 
     + \rd_{\beta 0}\phi_\alpha} 
  e^{-\rd_{\beta 0}S_\alpha(z)} 
  = - e^{-\rd_{\beta 0}S_\alpha(z)}.  
\eeqnn
Thus the foregoing equation boils down to 
\beqnn
  p_\alpha(z) - q_\beta 
  = e^{-\rd_{\beta 0}S_\alpha(z)}. 
\eeqnn
We can thereby recover (\ref{HJ-geneq4}).  
\qed

\section{Tau function of mutli-component KP hierarchy}

\subsection{Fermionic formula of tau function}

Following Date et al. \cite{DJKM-III} (see also 
the work of Kac and van de Leur \cite{KvdL-mcKP}), 
we now consider an $N+1$-component and 
``charged'' version of the KP hierarchy 
in their fermionic formalism.  
This hierarchy has an $N+1$-tuple $\bst$ of 
time variables organized in the vector notation as 
\beqnn
  \bst = (\bst_0,\bst_1,\ldots,\bst_N), \quad 
  \bst_\alpha = (t_{\alpha 1},t_{\alpha 2},\ldots), 
\eeqnn
and extra ``charge'' variables 
\beqnn
  \bss = (s_0,s_1,\ldots,s_N), \quad 
  \sum_{\alpha=0}^N s_\alpha = 0. 
\eeqnn
The charge variables correspond to 
the $t_{\alpha 0}$'s of the universal Whitham hierarchy.  

The fermionic formalism is based on 
an $N+1$-component free fermion system.  
The one-particle creation-annihilation operators of 
this system are labeled by $\alpha = 0,1,\ldots,N$ 
as $\psi_{\alpha j},\psi^*_{\alpha j}$ ($j \in \ZZ$), 
and obey the anticommuation relations 
\beq
  [\psi_{\alpha j},\psi^*_{\beta k}]_{+} 
  = \delta_{\alpha\beta}\delta_{jk}, \quad 
  [\psi_{\alpha j},\psi_{\beta k}]_{+} 
  = [\psi^*_{\alpha j},\psi^*_{\beta k}]_{+} = 0. 
\eeq
They form an infinite dimensional Clifford algebra.  
The Fock and dual Fock spaces are generated by 
the vacuum states $|\bszero\rangle$ 
and $\langle\bszero|$ that satisfy 
the annihilation conditions 
\beq
\begin{aligned}
  \langle\bszero|\psi_{\alpha j} = 0 \quad (j \ge 0), 
  &\qquad
  \psi_{\alpha j}|\bszero\rangle = 0 \quad (j < 0), \\
  \langle\bszero|\psi^*_{\alpha j} = 0 \quad (j < 0), 
  &\qquad 
  \psi^*_{\alpha j}|\bszero\rangle = 0 \quad (j \ge 0). 
\end{aligned}
\eeq
These spaces are decomposed to eigenspaces of 
the charge operators 
\beqnn
  H_{\alpha 0} 
  = \sum_{j=-\infty}^\infty :\psi_{\alpha j}\psi^*_{\alpha j}: 
  \quad (\text{normal ordering}), 
\eeqnn
each eigenspace being labeled by the charge vector 
$\bss = (s_0,s_1,\ldots,s_N)$ mentioned above.  
We can choose a ground state $|\bss\rangle$ and 
its dual $\langle\bss|$ in the charge $\bss$ sector as 
\beq
  \langle\bss| 
  = \langle\bszero|\Phi_{0s_0}\Phi_{1s_1}\cdots\Phi_{Ns_N}, 
  \quad 
  |\bss\rangle 
  = \Phi^*_{Ns_N}\cdots\Phi^*_{1s_1}\Phi^*_{0s_0}|\bszero\rangle, 
\eeq
where 
\beqnn
\begin{aligned}
  \Phi_{\alpha s_\alpha} 
  &= \begin{cases}
     \psi^*_{\alpha,0}\cdots\psi^*_{\alpha,s-1} 
     & (s > 0),\\
     \psi_{\alpha,-1}\cdots\psi_{\alpha,s} 
     & (s < 0),  
     \end{cases}
\\
  \Phi^*_{\alpha s_\alpha}
  &= \begin{cases}
     \psi_{\alpha,s-1}\cdots\psi_{\alpha,0}
     & (s > 0),\\
     \psi^*_{\alpha,s}\cdots\psi^*_{\alpha,-1}
     & (s < 0). 
     \end{cases}
\end{aligned}
\eeqnn
We then have the relation 
\beq
  \langle\bss|\psi_{\alpha,s_\alpha-1} 
  = \epsilon_\alpha(\bss)\langle \bss-\bse_\alpha |, 
  \quad 
  \langle\bss|\psi^*_{\alpha,s_\alpha-1} 
  = \epsilon_\alpha(\bss)\langle \bss+\bse_\alpha |  
\eeq
and a similar relation for $|\bss\rangle$, 
where $\bse_\alpha$ denotes the unit vector 
\beqnn
  \bse_\alpha = (\ldots,0,1,0,\ldots ) 
  \quad (\text{$1$ in the $\alpha$-th component}) 
\eeqnn
and $\epsilon_\alpha(\bss)$ the sign factor 
\beqnn
  \epsilon_\alpha(\bss) 
  = (-1)^{s_{\alpha+1}+\cdots+s_N}. 
\eeqnn

The $\tau$-function $\tau(\bss,\bst)$ of 
the $N+1$-component KP hierarchy is 
given by the expectation value 
\beq
  \tau(\bss,\bst) 
  = \langle\bss| e^{H(\bst)}g |\bszero\rangle 
\label{tau-def}
\eeq
of a product of two elements $e^{H(\bst)}$ 
and $g$ of the Clifford groups (i.e., 
the group of invertible elements of 
the Clifford algebra whose adjoint action 
generates a linear transformation on 
the linear span of one-particle 
creation-annihilation operators).  
$H(\bst)$ is the linear combination 
\beqnn
  H(\bst) 
  = \sum_{\alpha=0}^N\sum_{n=1}^\infty 
    t_{\alpha n}H_{\alpha n} 
\eeqnn
of part of the generators 
\beq
  H_{\alpha n} 
  = \sum_{j=-\infty}^\infty 
      :\psi_{\alpha,j}\psi^*_{\alpha,j+n}: 
\eeq
of an $N+1$-component Heisenberg algebra. 
$g$ is a general element of the Clifford group 
whose adjoint action on the linear span of 
one-particle operators does not mix  
$\psi_{\alpha j}$'s and $\psi^*_{\alpha j}$'s: 
\beq
  g\psi_{\beta k}g^{-1} 
  = \sum_{\alpha=0}^N \sum_{j=-\infty}^\infty 
    a_{\alpha\beta jk}\psi_{\alpha j}, \quad 
  g\psi^*_{\beta k}g^{-1}
  = \sum_{\alpha=0}^N \sum_{j=-\infty}^\infty 
    \tilde{a}_{\alpha\beta jk}\psi^*_{\alpha j}. 
\eeq
A typical case is the exponentiated 
fermion bilinear form 
\beqnn
  g = \exp\Bigl(\sum_{\alpha,\beta=0}^N 
        \sum_{j,k=-\infty}^\infty 
        a_{\alpha\beta jk}
        :\psi_{\alpha j}\psi^*_{\beta k}:\Bigr). 
\eeqnn
Note that $e^{H(\bst)}$, too, is an operator 
of this type.  

We have two important consequences of this setting. 
Firstly, the expectation value of (\ref{tau-def}) 
vanishes unless $s_0 + s_1 + \cdots + s_N = 0$.  
Secondly, the infinite matrices of the coefficients 
$a_{\alpha\beta jk}$ and $\tilde{a}_{\alpha\beta jk}$ 
turn out to be contragradient to each other 
(i.e., equal to the transposed inverse).  
This implies that the operator bilinear identity 
\beq
    \sum_{\gamma=0}^N\sum_{j=-\infty}^\infty 
    \psi_{\gamma j}g \otimes \psi^*_{\gamma j}g 
  = \sum_{\gamma=0}^N\sum_{j=-\infty}^\infty 
    g\psi_{\gamma j} \otimes g\psi^*_{\gamma j} 
\label{psi-g-bilineq}
\eeq
holds, and this identity leads to 
bilinear equations for the $\tau$-function.

\subsection{Bilinear equations of tau function}

Let us introduce the free fermion fields 
\beqnn
  \psi_\alpha(z) 
  = \sum_{j=-\infty}^\infty \psi_{\alpha j}z^j, \quad 
  \psi^*_\alpha(z) 
  = \sum_{j=-\infty}^\infty \psi^*_{\alpha j}z^{-j-1}. 
\eeqnn
and rewrite the operator identity (\ref{psi-g-bilineq}) 
to an integral form as 
\beq
    \sum_{\gamma=0}^N \oint\frac{dz}{2\pi i} 
    \psi_\gamma(z)g \otimes \psi^*_\gamma(z)g 
  = \sum_{\gamma=0}^N \oint\frac{dz}{2\pi i} 
    g\psi_\gamma(z) \otimes g\psi^*_\gamma(z). 
\eeq
The contour integral is understood to be 
an integral along the circle $|z| = R$ 
with sufficiently large radius $R$: 
\beqnn
  \oint\frac{dz}{2\pi i}z^n = \delta_{n,-1}. 
\eeqnn

We now apply both hand side of this operator identity 
to $|\bszero\rangle\otimes|\bszero\rangle$.  
A clue is the identity 
\beqnn
  \psi_{\gamma j}|\bszero\rangle 
  \otimes \psi^*_{\gamma j}|\bszero\rangle 
  = 0, 
\eeqnn
which holds for all $j \in \ZZ$ because 
either $\psi_{\gamma j}$ or $\psi^*_{\gamma j}$ 
annihilates $|\bszero\rangle$.  
This identity of states in the Fock space 
implies that 
\beqnn
    \sum_{\gamma=0}^N\oint\frac{dz}{2\pi i} 
      g\psi_\gamma(z)|\bszero\rangle \otimes 
      g\psi^*_\gamma(z)|\bszero\rangle
  = \sum_{\gamma=0}^N\sum_{j=-\infty}^\infty 
      g\psi_{\gamma j}|\bszero\rangle \otimes 
      g\psi^*_{\gamma j}|\bszero\rangle 
  = 0. 
\eeqnn
Thus we are left with the identity 
\beq
  \sum_{\gamma=0}^N 
    \oint\frac{dz}{2\pi i} 
    \psi_\gamma(z)g|\bszero\rangle 
    \otimes \psi^*_\gamma(z)g|\bszero\rangle = 0. 
\eeq

We now apply $e^{H(\bst')}\otimes e^{H(\bst)}$ 
to the left hand side of the last identity and 
make the inner product with $\langle\bss'+\bse_\alpha|
\otimes\langle\bss-\bse_\beta|$, where 
$\bst$, $\bst'$, $\bss$, $\bss'$, $\alpha$ 
and $\beta$ are arbitrary.  
This yields the equation 
\beq
  \sum_{\gamma=0}^N 
    \oint\frac{dz}{2\pi i} 
    \langle\bss'+\bse_\alpha|e^{H(\bst')}
      \psi_\gamma(z)g|\bszero\rangle 
    \langle\bss-\bse_\beta|e^{H(\bst)}
      \psi^*_\gamma(z)g|\bszero\rangle 
  = 0 
\eeq
of expectation values.  By a multi-component 
analogue of the well known bosonization formula,  
the expectation values in the integral 
can be expressed in terms of 
the tau function as 
\beqnn
\begin{aligned}
  \langle\bss'+\bse_\alpha|e^{H(\bst')}
    \psi_\gamma(z)g|\bszero\rangle 
  &= \epsilon_{\alpha\gamma}(\bss') 
     z^{s'_\gamma+\delta_{\alpha\gamma}-1} 
     e^{\xi(\bst'_\gamma,z)} 
     e^{-D_\gamma(z)}
     \tau(\bss'+\bse_\alpha-\bse_\gamma,\bst'), \\
  \langle\bss+\bse_\beta|e^{H(\bst)} 
    \psi^*_\gamma(z)g|\bszero\rangle 
  &= \epsilon_{\beta\gamma}(\bss) 
     z^{-s_\gamma+\delta_{\beta\gamma}-1}
     e^{-\xi(\bst_\gamma,z)} 
     e^{D_\gamma(z)} 
     \tau(\bss-\bse_\beta+\bse_\gamma,\bst), 
\end{aligned}
\eeqnn
where $D_\gamma(z)$ is the differential operator 
that has been used in the formula (\ref{S(z)-F}) 
of the $S$-functions, and the other notations 
are those commonly used for the (multi-component) 
KP hierarchy, namely, 
\beqnn
  \epsilon_{\beta\gamma}(\bss) 
  = \begin{cases}
    (-1)^{s_{\beta+1} + \cdots + s_\gamma} 
      & (\beta \le \gamma), \\
    - (-1)^{s_{\gamma+1} + \cdots + s_\beta}
      & (\beta > \gamma), 
    \end{cases}
  \quad
  \xi(\bst_\gamma,z) 
  = \sum_{n=1}^\infty t_{\gamma n}z^n. 
\eeqnn
Thus the foregoing bilinear equation 
of expectation values turns into 
the bilinear equation 
\beq
\begin{aligned}
  &\sum_{\gamma=0}^N 
    \epsilon_{\alpha\gamma}(\bss')
    \epsilon_{\beta\gamma}(\bss) 
  \oint\frac{dz}{2\pi i} 
    z^{s'_\gamma - s_\gamma + \delta_{\alpha\gamma} 
       + \delta_{\beta\gamma} - 2} 
    e^{\xi(\bst'_\gamma-\bst_\gamma,z)}
    \times \mbox{} \\
  &\mbox{} \times 
    (e^{-D_\gamma(z)}\tau)
      (\bss'+\bse_\alpha-\bse_\gamma,\,\bst') 
    (e^{D_\gamma(z)}\tau)
      (\bss-\bse_\beta+\bse_\gamma,\,\bst) 
  = 0 
\end{aligned}
\label{tau-bilineq}
\eeq
of the $\tau$-function, which holds 
for arbitrary value of $\bst'$,$\bst$, 
$\bss'$, $\bss$ and for 
$\alpha,\beta = 0,1,\ldots,N$.  
By the standard procedure, 
this bilinear equation can be converted to 
an infinite number of Hirota equations.  

The subsequent consideration is focused 
on the case where $\alpha=\beta=0$, 
which actually covers the general case 
by shifting $\bss$ and $\bss'$. 
For convenience, we separate the term of 
$\gamma = 0$ from the sum over $\gamma 
= 0,1,\ldots,N$ and move the sign factors 
$\epsilon_{00}(\bss')$ and 
$\epsilon_{00}(\bss)$ to the other terms.  
$\epsilon_{0\gamma}(\bss')$ and 
$\epsilon_{0\gamma}(\bss)$ of the latter 
are thereby replaced by 
\beqnn
  \tilde{\epsilon}_\gamma(\bss') 
  = \frac{\epsilon_{0\gamma}(\bss')}
         {\epsilon_{00}(\bss')} 
  = (-1)^{s'_1+\cdots+s'_\gamma}, 
  \quad 
  \tilde{\epsilon}_\gamma(\bss) 
  = \frac{\epsilon_{0\gamma}(\bss)}
         {\epsilon_{00}(\bss)} 
  = (-1)^{s_1+\cdots+s_\gamma}. 
\eeqnn
Bilinear equation (\ref{tau-bilineq}) 
for $\alpha = \beta = 0$ thus reads 
\beq
\begin{aligned}
  &\oint\frac{dz}{2\pi i}
   z^{s'_0-s_0}e^{\xi(\bst'_0-\bst_0,z)} 
   (e^{-D_0(z)}\tau)(\bss',\bst')
   (e^{D_0(z)}\tau)(\bss,\bst)\\ 
  \mbox{} 
  + &\sum_{\gamma=1}^N 
    \tilde{\epsilon}_\gamma(\bss')
    \tilde{\epsilon}_\gamma(\bss) 
  \oint\frac{dz}{2\pi i}
    z^{s'_\gamma-s_\gamma-2} 
    e^{\xi(\bst'_\gamma-\bst_\gamma,z)} 
    \times \mbox{}\\
  &\mbox{} \times 
    (e^{-D_\gamma(z)}\tau)(\bss'+\bse_0-\bse_\gamma,\,\bst')  
    (e^{D_\gamma(z)}\tau)(\bss-\bse_0+\bse_\gamma,\,\bst) 
  = 0. 
\end{aligned}
\label{tau-bilineq00}
\eeq
We shall show later that the sign factor 
$\tilde{\epsilon}_\gamma(\bse)$ is related 
to the factor proportional to $\log(-1)$ 
in the defining equations (\ref{F-def}) of 
the $F$-function.

\section{Dispersionless Hirota equations as 
dipsersionless limit of differential Fay identities}

\subsection{How to derive differential Fay identity}

Let us recall the differential Fay identity 
for the $\tau$-function $\tau(\bst)$, $\bst 
= (t_1,t_2,\ldots)$, of the KP hierarchy \cite{AvM-92}.  
This identity is usually derived by taking 
a limit of the three-term Fay identity 
\beq
\begin{aligned}
& (z_0 - z_1)(z_2 - z_3)
  \tau(\bst+[z_0]+[z_1])\tau(\bst+[z_2]+[z_3])\\
& \mbox{}
  + (\text{cyclic permutations of $z_1,z_2,z_3$}) 
  = 0, 
\end{aligned}
\eeq
where $[z]$ denotes the shift vector 
\beqnn
  [z] = (z,z^2/2,\ldots,z^n/n,\ldots) 
\eeqnn
in the time variables.  

Actually, there is an alternative method 
to derive the differential Fay identity 
as follows.  This alternative method consists 
of the following steps. The first step is 
to differentiate the bilinear equation 
\beq
  \oint\frac{dz}{2\pi i}e^{\xi(\bst'-\bst,z)} 
    (e^{-D(z)}\tau)(\bst')(e^{D(z)}\tau)(\bst) 
  = 0 
\eeq
by $t'_1$.  This yields an equation of the form 
\beqnn
  \oint\frac{dz}{2\pi i}e^{\xi(\bst'-\bst,z)} 
    \bigl(z(e^{-D(z)}\tau)(\bst') 
     + (\rd_1e^{-D(z)}\tau)(\bst')\bigr) 
    \tau(\bst+[z^{-1}]) 
  = 0, 
\eeqnn
where $\rd_1$ denotes the differential operator 
\beqnn
  \rd_1 = \rd/\rd t_1. 
\eeqnn
The next step is to specialize $\bst'$ as 
\beqnn
  \bst' = \bst + [\lambda^{-1}] + [\mu^{-1}], 
\eeqnn
where $\lambda$ and $\mu$ are arbitrary constants 
that sit on the far side of the contour $|z| = R$ 
of the integral, namely, 
\beqnn
  |\lambda| > R, \quad |\mu| > R. 
\eeqnn
The exponential factor $e^{\xi(\bst'-\bst,z)}$ 
thereby reduces to 
\beqnn
  e^{\xi(\bst'-\bst,z)} 
  = \Bigl(1 - \frac{z}{\lambda}\Bigr)^{-1} 
    \Bigl(1 - \frac{z}{\mu}\Bigr)^{-1} 
  = \frac{\lambda\mu}{(z-\lambda)(z-\mu)},  
\eeqnn
and the foregoing bilinear equation 
now takes such a form as 
\beqnn
\begin{aligned}
  \oint
   \frac{dz}{2\pi i}
   \frac{\lambda\mu}{(z-\lambda)(z-\mu)} 
  &\bigl(z\tau(\bst+[\lambda^{-1}]+[\mu^{-1}]-[z^{-1}]) 
     + \mbox{} \\
  &\mbox{} 
     + (\rd_1\tau)(\bst+[\lambda^{-1}]+[\mu^{-1}]-[z^{-1}])
   \bigr)\tau(\bst+[z^{-1}]) = 0. 
\end{aligned}
\eeqnn
Assuming, as usual, that the factors including 
the $\tau$-function are holomorphic functions 
of $z$ on the far side of the contour, 
one can deform the contour to a union of 
circles that encircle the poles at 
$z = \lambda,\mu,\infty$.  The contour integrals 
along those circles are given by residues.  
Collecting these pieces, one eventually obtains 
the equation 
\beq
\begin{aligned}
& (\lambda-\mu)\tau(\bst+[\lambda^{-1}]+[\mu^{-1}])\tau(\bst) 
  - \tau(\bst+[\lambda^{-1}])\tau(\bst+[\mu^{-1}]) \\
& \mbox{} 
  + (\rd_1\tau)(\bst+[\lambda^{-1}])
    \tau(\bst+[\mu^{-1}]) 
  - (\rd_1\tau)(\bst+[\mu^{-1}]) 
    \tau(\bst+[\lambda^{-1}]) 
  = 0, 
\end{aligned}
\eeq
which is essentially the differential Fay identity. 

This equation can be cast into a form that is 
more suited for considering the dispersionless limit.  
Let us divide both hand side of this equation 
by $(\lambda-\mu)\tau(\bst+[\lambda^{-1}])
\tau(\bst+[\mu^{-1}])$.  This leads to 
an equation of the form 
\beq
  \frac{\tau(\bst+[\lambda^{-1}]+[\mu^{-1}])\tau(\bst)}
      {\tau(\bst+[\lambda^{-1}])\tau(\bst+[\mu^{-1}])} 
  = 1 
    - \frac{\rd_1\bigl(\log\tau(\bst+[\lambda^{-1}]) 
            - \log\tau(\bst+[\mu^{-1}])\bigr)} 
           {\lambda - \mu}. 
\eeq
Using the differential operator 
\beqnn
  D(z) = \sum_{n=1}^\infty \frac{z^{-n}}{n}\rd_n, 
  \quad \rd_n = \rd/\rd t_n, 
\eeqnn
we can rewrite the last equation as 
\beq
  \exp\bigl((e^{D(\lambda)}-1)(e^{D(\mu)}-1)
            \log\tau(\bst)\bigr) 
  = 1 - \frac{\rd_1(e^{D(\lambda)}-e^{D(\mu)})
              \log\tau(\bst)}{\lambda-\mu}.  
\eeq
This shows a fully ``dispersive'' form of 
the dispersionless Hirota equation 
\beq
  e^{D(\lambda)D(\mu)F} 
  = 1 - \frac{\rd_1(D(\lambda)-D(\mu))F}{\lambda-\mu}. 
\eeq

\subsection{Multi-component analogue of 
differential Fay identity}

The foregoing method turns out to work for 
the bilinear equations (\ref{tau-bilineq00}) 
of the multi-component case as well.  
We thus obtain a multi-component analogue, 
(\ref{dFay-eq1})--(\ref{dFay-eq4}), of 
the differential Fay identity as follows. 

\begin{theorem}
The $\tau$-function satisfies the multi-component 
differential Fay identities 
\beq
\lefteqn{
  \exp\bigl((e^{D_0(\lambda)}-1)(e^{D_0(\mu)}-1) 
      \log\tau(\bss,\bst)\bigr) 
}\nonumber\\
&&\qquad\qquad\qquad
= 1 
  - \frac{\rd_{01}\bigl(e^{D_0(\lambda)}-e^{D_0(\mu)}
      \bigr)\log\tau(\bss,\bst)}{\lambda-\mu}, 
\label{dFay-eq1}\\
\lefteqn{
  \lambda\exp\bigl((e^{D_0(\lambda)}-1)
   (e^{\rd_{\alpha 0}+D_\alpha(\mu)}-1)
   \log\tau(\bss,\bst)\bigr) 
}\nonumber\\
&&\qquad\qquad\qquad
= \lambda 
  - \rd_{01}(e^{D_0(\lambda)}
    - e^{\rd_{\alpha 0}+D_\alpha(\mu)})
    \log\tau(\bss,\bst), 
\label{dFay-eq2}\\
\lefteqn{
  \exp\bigl((e^{\rd_{\alpha 0}+D_\alpha(\lambda)}-1) 
            (e^{\rd_{\alpha 0}+D_\alpha(\mu)}-1)
            \log\tau(\bss,\bst) \bigr) 
}\nonumber\\
&&\qquad\qquad\qquad
= -\frac{\lambda\mu\rd_{01}
     \bigl(e^{\rd_{\alpha 0}+D_\alpha(\lambda)}
         - e^{\rd_{\alpha 0}+D_\alpha(\mu)}\bigr)
     \log\tau(\bss,\bst)}{\lambda-\mu}, 
\label{dFay-eq3}\\
\lefteqn{
  \epsilon_{\alpha\beta}
  \exp\bigl((e^{\rd_{\alpha 0}+D_\alpha(\lambda)}-1) 
    (e^{\rd_{\beta 0}+D_\beta(\mu)}-1)
    \log\tau(\bss,\bst) \bigr) 
}\nonumber\\
&&\qquad\qquad\qquad
= - \rd_{01}
    \bigl(e^{\rd_{\alpha 0}+D_\alpha(\lambda)}
        - e^{\rd_{\beta 0}+D_\beta(\mu)}\bigr)
    \log\tau(\bss,\bst), 
\label{dFay-eq4}
\eeq
where $e^{\rd_{\alpha 0}}$ is understood to be 
the shift operator in the $\bss$ variables, i.e., 
\beqnn
  e^{\rd_{\alpha 0}}f(\bss) 
  = f(\bss-\bse_0+\bse_\alpha). 
\eeqnn
\end{theorem}
  
\proof
To derive (\ref{dFay-eq1}), we differentiate 
(\ref{tau-bilineq00}) by $t'_{01}$ and put 
\beqnn
  \bst'_0 = \bst_0 + [\lambda^{-1}] + [\mu^{-1}], 
  \quad 
  \bst'_\alpha = \bst_\alpha \quad (\alpha=1,\ldots,N), 
  \quad 
  \bss' = \bss. 
\eeqnn
The differentiated bilinear equation 
thereby simplifies as 
\beqnn
\begin{aligned}
  \oint
   &\frac{dz}{2\pi i}
   \frac{\lambda\mu}{(z-\lambda)(z-\mu)} 
   \bigl(z\tau(\bss,\,\bst+[\lambda^{-1}]_0
              +[\mu^{-1}]_0-[z^{-1}]_0) 
    + \mbox{} \nonumber\\
   & \mbox{} 
    + (\rd_{01}\tau)(\bss,\,\bst+[\lambda^{-1}]_0 
                +[\mu^{-1}]_0-[z^{-1}]_0) 
   \bigr)\tau(\bss,\,\bst+[z^{-1}]_0) 
  = 0, 
\end{aligned}
\eeqnn
where $[z]_\alpha$ now stands for the shift vector 
\beqnn
  [z]_\alpha = (\ldots,\bszero,[z],\bszero,\ldots) 
  \quad (\text{$[z]$ in the $\alpha$-th component}) 
\eeqnn
in the multi-component time variables.  
This is substantially the same bilinear equation 
as mentioned above for the case of the usual 
KP hierarchy.  We thus obtain the equation 
\beqnn
\begin{aligned}
& \frac{\tau(\bss,\,\bst+[\lambda^{-1}]_0+[\mu^{-1}]_0) 
        \tau(\bss,\bst)} 
       {\tau(\bss,\,\bst+[\lambda^{-1}]_0) 
        \tau(\bss,\,\bst+[\mu^{-1}]_0)} \\
&= 1 
 - \frac{\rd_{01}\bigl(\log\tau(\bss,\,\bst+[\lambda^{-1}]_0)
         - \log\tau(\bss,\,\bst+[\mu^{-1}]_0)\bigr)}
        {\lambda-\mu}, 
\end{aligned}
\eeqnn
which we can readily rewrite to 
(\ref{dFay-eq1}).  

To derive (\ref{dFay-eq2}), we differentiate 
(\ref{tau-bilineq00}) by $t'_{01}$ and put 
\beqnn
  \bst'_0 = \bst_0 + [\lambda^{-1}], \quad 
  \bst'_\alpha = \bst_\alpha + [\mu^{-1}], \quad 
  \bst'_\beta = \bst_\beta \quad (\beta\not=\alpha), \\
  s'_0 = s_0 - 1, \quad 
  s'_\alpha = s_\alpha + 1, \quad 
  s'_\beta = s_\beta \quad (\beta\not=\alpha). 
\eeqnn
The differentiated bilinear equation takes 
such a form as 
\beqnn
\begin{aligned}
- \oint
    &\frac{dz}{2\pi i}
     \frac{1}{z}\frac{-\lambda}{z-\lambda} 
     \bigl(z\tau(\bss-\bse_0+\bse_\alpha,\,
      \bst+[\lambda^{-1}]_0-[z^{-1}]_0 
      +[\mu^{-1}]_\alpha) + \mbox{}\\
    &\quad\mbox{} 
      + (\rd_{01}\tau)(\bss-\bse_0+\bse_\alpha,\,
          \bst+[\lambda^{-1}]_0-[z^{-1}]_0 
          +[\mu^{-1}]_\alpha) \bigr)\times \mbox{}\\
    &\mbox{} \times
     \tau(\bss,\,\bst+[z^{-1}]_0) \\
+ \oint
    &\frac{dz}{2\pi i} 
     \frac{1}{z}\frac{-\mu}{z-\mu} 
     (\rd_{01}\tau)(\bss,\,\bst+[\lambda^{-1}]_0
       +[\mu^{-1}]_\alpha-[z^{-1}]_\alpha) \times \mbox{}\\
    &\mbox{} 
     \times 
     \tau(\bss-\bse_0+\bse_\alpha,\,\bst+[z^{-1}]_\alpha) 
= 0. 
\end{aligned}
\eeqnn
After calculating the contour integrals, 
this equation reduces to 
\beqnn
\begin{aligned}
 & \lambda\tau(\bss-\bse_0+\bse_\alpha,\,
     \bst+[\lambda^{-1}]_0+[\mu^{-1}]_\alpha)\tau(\bss,\bst) \\
 & - \lambda\tau(\bss-\bse_0+\bse_\alpha,\,\bst+[\mu^{-1}]_\alpha) 
     \tau(\bss,\,\bst+[\lambda^{-1}]_0) \\
 & - (\rd_{01}\tau)(\bss-\bse_0+\bse_\alpha,\,\bst+[\mu^{-1}]_\alpha) 
     \tau(\bss,\,\bst+[\lambda^{-1}]_0) \\
 & + (\rd_{01}\tau)(\bss,\,\bst+[\lambda^{-1}]_0) 
      \tau(\bss-\bse_0+\bse_\alpha,\,\bst+[\mu^{-1}]_\alpha) 
  = 0, 
\end{aligned}
\eeqnn
which we can further rewrite as 
\beqnn
\begin{aligned}
 & \lambda 
   \frac{\tau(\bss-\bse_0+\bse_\alpha,\,
         \bst+[\lambda^{-1}]_0+[\mu^{-1}]_\alpha) 
         \tau(\bss,\bst)} 
        {\tau(\bss,\,\bst+[\lambda^{-1}]_0) 
         \tau(\bss,\,\bst+[\mu^{-1}]_\alpha)} \\
 &= \lambda 
    - \rd_{01}\bigl(\log\tau(\bss,\,\bst+[\lambda^{-1}]_0)
        - \log\tau(\bss-\bse_0+\bse_\alpha,\,
          \bst+[\mu^{-1}]_\alpha) \bigr). 
\end{aligned}
\eeqnn
After some more algebra, this equation 
becomes (\ref{dFay-eq2}).  

To derive (\ref{dFay-eq3}), we choose 
\beqnn
  \bst'_\alpha = \bst_\alpha + [\lambda^{-1}] + [\mu^{-1}], 
  \quad
  \bst'_\beta = \bst_\beta \quad (\beta \not= \alpha), \\
  s'_0 = s_0 -2, \quad 
  s'_\alpha = s_\alpha + 2, \quad 
  s'_\beta = s_\beta \quad (\beta \not= 0,\alpha). 
\eeqnn
The differentiated bilinear equation now reads 
\beqnn
\begin{aligned}
  \oint 
  &\frac{dz}{2\pi i}
   \bigl(z\tau(\bss-2\bse_0+2\bse_\alpha,\,\bst-[z^{-1}]_0
               +[\lambda^{-1}]_\alpha+[\mu^{-1}]_\alpha) 
      + \mbox{} \\
  &\quad\mbox{} 
      + (\rd_{01}\tau)(\bss-\bse_0+\bse_\alpha,\,
         \bst-[z^{-1}]_0 +[\lambda^{-1}]_\alpha+[\mu^{-1}]_\alpha) 
    \bigr) \times \mbox{} \\
  & \mbox{} \times \tau(\bss,\,\bst+[z^{-1}]_0) \\
+ \oint
  &\frac{dz}{2\pi i}
     \frac{\lambda\mu}{(z-\lambda)(z-\mu)} 
     (\rd_{01}\tau)(\bss-\bse_0+\bse_\alpha,\,
        \bst+[\lambda^{-1}]_\alpha+[\mu^{-1}]_\alpha) 
   \times \mbox{}\\
  & \mbox{} \times 
     \tau(\bss-\bse_0+\bse_\alpha,\,\bst+[z^{-1}]_0) 
  = 0. 
\end{aligned}
\eeqnn 
The outcome of residue calculus is the equation 
\beqnn
\begin{aligned}
  &\frac{\tau(\bss-2\bse_0+2\bse_\alpha,\,\bst+[\lambda^{-1}]_\alpha
              +[\mu^{-1}]_\alpha)\tau(\bss,\bst)} 
        {\tau(\bss-\bse_0+\bse_\alpha,\,\bst+[\lambda^{-1}]_\alpha)
         \tau(\bss-\bse_0+\bse_\alpha,\,\bst+[\mu^{-1}]_\alpha)} 
  = - \frac{\lambda\mu}{\lambda-\mu} \times \mbox{} \\
  &\mbox{} \times 
     \rd_{01}\bigl(
       \log\tau(\bss-\bse_0+\bse_\alpha,\,\bst+[\lambda^{-1}]_\alpha) 
       - \log\tau(\bss-\bse_0+\bse_\alpha,\,\bst+[\mu^{-1}]_\alpha) 
     \bigr), 
\end{aligned}
\eeqnn
which becomes (\ref{dFay-eq3}) after some algebra. 

Lastly, we derive (\ref{dFay-eq4}). 
This is achieved by choosing 
\beqnn
  \bst'_\alpha = \bst_\alpha + [\lambda^{-1}], \quad 
  \bst'_\beta = \bst_\beta + [\mu^{-1}], \quad 
  \bst'_\gamma = \bst_\gamma \quad 
    (\gamma \not= \alpha,\beta), \\
  s'_0 = s_0 - 2, \quad 
  s'_\alpha = s_\alpha + 1, \quad 
  s'_\beta = s_\beta + 1, \quad 
  s'_\gamma = s_\gamma \quad 
    (\gamma \not= \alpha,\beta), 
\eeqnn
and repeating the same procedure as 
the previous cases.  Unlike those cases, 
however, we now have to account for 
the sign factors in (\ref{tau-bilineq00}). 
The outcome takes a different form for 
$\alpha < \beta$ and $\alpha > \beta$ as follows: 
\beqnn
\begin{aligned}
 \oint 
  &\frac{dz}{2\pi i}\frac{1}{z^2} 
   \bigl(z\tau(\bss-2\bse_0+\bse_\alpha+\bse_\beta,\, 
           \bst-[z^{-1}]_0+[\lambda^{-1}]_\alpha
           +[\mu^{-1}]_\beta) + \mbox{}\\
  &\quad\mbox{} 
       + (\rd_{01}\tau)(\bss-2\bse_0+\bse_\alpha+\bse_\beta,\, 
           \bst-[z^{-1}]_0+[\lambda^{-1}]_\alpha
           +[\mu^{-1}]_\beta)\bigr) \times \mbox{}\\ 
  &\mbox{} \times 
   \tau(\bss,\,\bst+[z^{-1}]_0) \\
-\epsilon_{\alpha\beta}\oint 
  &\frac{dz}{2\pi i}\frac{1}{z}\frac{-\lambda}{z-\lambda} 
   (\rd_{01}\tau)(\bss-\bse_0+\bse_\beta,\, 
        \bst+[\lambda^{-1}]_\alpha-[z^{-1}]_\alpha
        +[\mu^{-1}]_\beta) \times \mbox{} \\
  &\mbox{} \times 
   \tau(\bss-\bse_0+\bse_\alpha,\,\bst+[z^{-1}]_\alpha) \\
+\epsilon_{\alpha\beta}\oint 
  &\frac{dz}{2\pi i}\frac{1}{z}\frac{-\mu}{z-\mu} 
   (\rd_{01}\tau)(\bse+\bse_\alpha,\, 
        \bst+[\lambda^{-1}]_\alpha+[\mu^{-1}]_\beta 
        -[z^{-1}]_\beta) \times \mbox{} \\
  &\mbox{} \times
   \tau(\bss-\bse_0+\bse_\beta,\,\bst+[z^{-1}]_\beta) 
  = 0, 
\end{aligned}
\eeqnn
where $\epsilon_{\alpha\beta}$ stands for 
the same sign factor as in (\ref{dHirota-eq4}).  
This equation reduces to 
\beqnn
\begin{aligned}
& \epsilon_{\alpha\beta}
  \frac{\tau(\bss-2\bse_0+\bse_\alpha+\bse_\beta,\, 
             \bst+[\lambda^{-1}]_\alpha+[\mu^{-1}]_\beta) 
        \tau(\bss,\bst)} 
       {\tau(\bss-\bse_0+\bse_\alpha,\,\bst+[\lambda^{-1}]_\alpha) 
        \tau(\bss-\bse_0+\bse_\beta,\,\bst+[\mu^{-1}]_\beta)} \\
&= - \rd_{01}\bigl( 
      \log\tau(\bss-\bse_0+\bse_\alpha,\,\bst+[\lambda^{-1}]_\alpha) 
      - \log\tau(\bss-\bse_0+\bse_\beta,\,\bst+[\mu^{-1}]_\beta) 
    \bigr), 
\end{aligned}
\eeqnn
and we eventually obtain (\ref{dFay-eq4}). 
\qed

\subsection{Dispersionless limit}

Dispersionless limit is realized as 
quasi-classical limit.  Namely, 
we allow the $\tau$-function to depend 
on the ``Planck constant'' $\hbar$ as well, 
i.e., $\tau = \tau(\hbar,\bss,\bst)$, 
and assume that the rescaled $\tau$-function 
\beqnn
  \tau_\hbar(\bss,\bst) 
  = \tau(\hbar,\hbar^{-1}\bss,\hbar^{-1}\bst) 
\eeqnn
behaves as 
\beq
  \tau_\hbar(\bss,\bst)  
  = \exp\bigl(\hbar^{-2}F(\bss,\bst) + O(\hbar^{-1})\bigr) 
\eeq
in the classical limit $\hbar \to 0$.  

The rescaled $\tau$-function satisfies 
(\ref{dFay-eq1})--(\ref{dFay-eq4}) 
with the operators $e^{D_0(\lambda)}$ etc. 
being also recalled: 
\beqnn
\lefteqn{
  \exp\bigl((e^{\hbar D_0(\lambda)}-1)(e^{\hbar D_0(\mu)}-1) 
      \log\tau_\hbar(\bss,\bst)\bigr) 
}\nonumber\\
&&\qquad\qquad\qquad
= 1 
  - \frac{\hbar\rd_{01}
      \bigl(e^{\hbar D_0(\lambda)}-e^{\hbar D_0(\mu)}\bigr)
      \log\tau_\hbar(\bss,\bst)}{\lambda-\mu}, 
\\
\lefteqn{
  \lambda\exp\bigl((e^{\hbar D_0(\lambda)-1}-1)
   (e^{\rd_{\hbar\alpha 0}+\hbar D_\alpha(\mu)}-1)
   \log\tau_\hbar(\bss,\bst)\bigr) 
}\nonumber\\
&&\qquad\qquad\qquad
= \lambda 
   - \hbar\rd_{01}
     \bigl(e^{D_0(\lambda)} 
      - e^{\hbar\rd_{\alpha 0}+\hbar D_\alpha(\mu)}\bigr) 
     \log\tau(\bss,\bst), 
\\
\lefteqn{
  \exp\bigl((e^{\hbar\rd_{\alpha 0}+\hbar D_\alpha(\lambda)}-1) 
            (e^{\hbar\rd_{\alpha 0}+\hbar D_\alpha(\mu)}-1)
            \log\tau_\hbar(\bss,\bst) \bigr) 
}\nonumber\\
&&\qquad\qquad\qquad
= - \frac{\lambda\mu\hbar\rd_{01}
      \bigl(e^{\hbar\rd_{\alpha 0}+\hbar D_\alpha(\lambda)} 
      - e^{\hbar\rd_{\alpha 0}+\hbar D_\alpha(\mu)}\bigr)
      \log\tau_\hbar(\bss,\bst)}{\lambda-\mu}, 
\\
\lefteqn{
  \epsilon_{\alpha\beta}
  \exp\bigl((e^{\hbar\rd_{\alpha 0}+\hbar D_\alpha(\lambda)}-1) 
    (e^{\hbar\rd_{\beta 0}+\hbar D_\beta(\mu)}-1)
    \log\tau_\hbar(\bss,\bst) \bigr) 
}\nonumber\\
&&\qquad\qquad\qquad
= - \hbar\rd_{01}
    \bigl(e^{\hbar\rd_{\alpha 0}+\hbar D_\alpha(\lambda)}
     - e^{\hbar\rd_{\beta 0}+\hbar D_\beta(\mu)}\bigr) 
    \log\tau_\hbar(\bss,\bst). 
\eeqnn
If we substitute the aforementioned 
quasi-classical ansatz of 
the rescaled $\tau$-function 
and take the limit as $\hbar \to 0$, 
we end up with the dipsersionless 
Hirota equations (\ref{dHirota-eq1})--(\ref{dHirota-eq4}) 
for the $F$-function.  In other words, 
the multi-component analogues 
(\ref{dFay-eq1})--(\ref{dFay-eq4}) 
of the differential Fay identity 
are ``dispersive'' counterparts of 
the dispersionless Hirota equations.

\section{Differential Fay identities as generating 
functional form of auxiliary linear equations}

\subsection{How to derive auxiliary linear equations}

It is known that the differential Fay identity 
is actually equivalent to the KP hierarchy itself. 
This is a consequence of the fact that 
the auxiliary linear equations of the KP hierarchy 
can be derived from the differential Fay identity.  
Let us briefly review this fact 
\cite[Appendix C]{TT-review} (see also Teo's paper 
\cite{Teo-06} for an analogous result and its proof 
for the Toda hierarchy). 

To derive auxiliary linear equations, 
we write the differential Fay identity as 
\beqnn
  \frac{\tau(\bst+[\lambda^{-1}]+[\mu^{-1}])\tau(\bst)} 
       {\tau(\bst+[\lambda^{-1}])\tau(\bst+[\mu^{-1}])} 
  = 1 
    + \frac{1}{\lambda-\mu}\rd_1\log
      \frac{\tau(\bst+[\mu^{-1}])}
           {\tau(\bst+[\lambda^{-1}])} 
\eeqnn
and shift the variables as $\bst \to 
\bst - [\lambda^{-1}] - [\mu^{-1}]$.  This leads to 
another form 
\beqnn
  \frac{\tau(\bst-[\lambda^{-1}]-[\mu^{-1}])\tau(\bst)} 
       {\tau(\bst-[\lambda^{-1}])\tau(\bst-[\mu^{-1}])} 
  = 1 
    + \frac{1}{\lambda-\mu}\rd_1\log
      \frac{\tau(\bst-[\lambda^{-1}])}
           {\tau(\bst-[\mu^{-1}])} 
\eeqnn
of the differential Fay identity.  
We can rewrite it as 
\beqnn
  e^{-D(\lambda)}\frac{\tau(\bst-[\mu^{-1}])}{\tau(\bst)} 
  = \frac{\tau(\bst-[\mu^{-1}])}{\tau(\bst)} 
    \Bigl(1 + \frac{1}{\lambda-\mu}\rd_1\log 
              \frac{\tau(\bst-[\lambda^{-1}])}
                   {\tau(\bst-[\mu^{-1}])} \Bigr), 
\eeqnn
which turns into the equation 
\beq
  \lambda e^{-D(\lambda)}\Psi(\bst,\mu) 
  = - \bigl(\rd_1 - \rd_1\log\Psi(\bst,\lambda)
      \bigr)\Psi(\bst,\mu) 
\label{KP-gen-lineq}
\eeq
for the wave function 
\beq
  \Psi(\bst,z) 
  = \frac{\tau(\bst-[z^{-1}])}{\tau(\bst)}e^{\xi(\bst,z)}. 
\eeq

(\ref{KP-gen-lineq}) may be thought of 
as a generating functional form of 
auxiliary linear equations for $\Psi(\bst,z)$.  
Let us introduce the fundamental Schur functions 
$h_j(\bst)$, $j = 0,1,2,\ldots$, 
by the generating function 
\beqnn
  e^{\xi(\bst,z)} = \sum_{j=0}^\infty h_j(\bst)z^j. 
\eeqnn
(\ref{KP-gen-lineq}) can be thereby 
decomposed to the linear equations 
\beq
  h_j(- \tilde{\rd}_{\bst})\Psi(\bst,\mu) 
  = f_j(\bst)\Psi(\bst,\mu) 
  \quad (j = 2,3,\ldots), 
\eeq
where $\tilde{\rd}_{\bst}$ denotes the vector
\beqnn
  \tilde{\rd}_{\bst} 
  = \Bigl(\rd_1,\frac{1}{2}\rd_2,\cdots, 
          \frac{1}{n}\rd_n,\ldots\Bigr) 
\eeqnn
of derivative operators, and $f_j(\bst)$'s are 
the coefficients of Laurent expansion of 
$\rd_1\log\Psi(\bst,z)$ at $z = \infty$, i.e., 
\beqnn
  \rd_1\log\Psi(\bst,z) 
  = z + \rd_1(e^{-D(z)}-1)\log\tau(\bst) 
  = z + \sum_{j=1}^\infty f_{j+1}(\bst)z^{-j}. 
\eeqnn
It is well known that these Laurent coefficients 
give conserved densities of the KP hierarchy.  

We can convert (\ref{KP-gen-lineq}) to linear equations 
of the familiar {\em evolutionary} form 
\beq
  \rd_n\Psi(\bst,z) = B_n\Psi(\bst,z) 
  \quad (n = 1,2,\ldots), 
\label{KP-lineq}
\eeq
where $B_n$ is a differential operator 
of the form 
\beqnn
  B_n = \rd_1^n + b_{n2}\rd_1^{n-2} + \cdots + b_{nn}. 
\eeqnn
This is achieved by the following method 
\cite[Appendix C]{TT-review}.  
Let us rewrite (\ref{KP-gen-lineq}) as 
\beqnn
  (1 - e^{-D(\lambda)})\Psi(\bst,\mu) 
  = X(\bst,\lambda)\Psi(\bst,\mu), 
\eeqnn
where $X(\lambda)$ denotes the differential operator 
\beqnn
  X(\bst,\lambda) 
  = \lambda^{-1}\rd_1 + 1 - \lambda^{-1}\rd_1\log\Psi(\bst,\lambda) 
\eeqnn
that depends on $\lambda$.  Let $X_k(\bst,\lambda)$, 
$k = 2,3,\ldots$, be differential operators 
defined by the recurrence relation 
\beqnn
  X_{k+1}(\bst,\lambda) 
  = X_k(\bst,\lambda) 
    - X_k(\bst-[\lambda^{-1}],\lambda)(1 - X_k(\bst,\lambda)), \quad
  X_1(\bst,\lambda) = X(\bst,\lambda).  
\eeqnn
$X_k(\bst,\lambda)$ is a differential operator of the form  
\beqnn
  X_k(\bst,\lambda) = \lambda^{-k}\rd_1^k + \cdots, 
\eeqnn
and turns out to satisfy the equation 
\beqnn
  (1 - e^{-D(\lambda)})^k\Psi(\bst,\mu) 
  = X_k(\bst,\lambda)\Psi(\bst,\mu).  
\eeqnn
Moreover, by construction, 
\beqnn
  X_k(\lambda) = O(\lambda^{-k}).  
\eeqnn
Consequently, the action of 
\beqnn
  D(\lambda) = - \log\bigl(1 - (1 - e^{-D(\lambda)})\bigr) 
  = \sum_{k=1}^\infty \frac{1}{k}(1 - e^{-D(\lambda)})^k 
\eeqnn
on $\Psi(\bst,\mu)$ is well defined, and we have 
the linear equation
\beqnn
  D(\lambda)\Psi(\bst,\mu) 
  = \sum_{k=1}^\infty \frac{1}{k}X_k(\bst,\lambda)\Psi(\bst,\mu), 
\eeqnn
which, upon expanded in powers of $\lambda$, 
gives the usual auxiliary linear equations 
(\ref{KP-lineq}) as expected.

\subsection{Generating functional form of 
auxiliary linear equation for multi-component case}

We now turn to the the multi-component case.   
Let us recall the following form of 
the differential Fay identities 
first derived from the bilinear equations 
(\ref{tau-bilineq00}): 
\beq
\lefteqn{
\frac{\tau(\bss,\,\bst+[\lambda^{-1}]_0+[\mu^{-1}]_0)\tau(\bss,\bst)} 
       {\tau(\bss,\,\bst+[\lambda^{-1}]_0)\tau(\bss,\bst+[\mu^{-1}]_0)} 
} \nonumber
\\
&&\qquad\qquad\qquad
= 1 + \frac{1}{\lambda-\mu}\rd_{01}\log
        \frac{\tau(\bss,\,\bst+[\mu^{-1}]_0)}
             {\tau(\bss,\,\bst+[\lambda^{-1}]_0)}, 
\label{dF-eq1}\\
\lefteqn{
\lambda
\frac{\tau(\bss-\bse_0+\bse_\alpha,\,
      \bst+[\lambda^{-1}]_0+[\mu^{-1}]_\alpha)\tau(\bss,\bst)}
     {\tau(\bss,\,\bst+[\lambda^{-1}]_0)
      \tau(\bss-\bse_0+\bse_\alpha,\,\bst+[\mu^{-1}]_\alpha)} 
} \nonumber\\
&&\qquad\qquad\qquad
= \lambda 
  + \rd_{01}\log
    \frac{\tau(\bss-\bse_0+\bse_\alpha,\,\bst+[\mu^{-1}]_\alpha)}
         {\tau(\bss,\,\bst+[\lambda^{-1}]_0)}, 
\label{dF-eq2}\\
\lefteqn{
\frac{\tau(\bss-2\bse_0+2\bse_\alpha,\,
      \bst+[\lambda^{-1}]_\alpha+[\mu^{-1}]_\alpha)\tau(\bss,\bst)}
     {\tau(\bss-\bse_0+\bse_\alpha,\,\bst+[\lambda^{-1}]_\alpha)
      \tau(\bss-\bse_0+\bse_\alpha,\,\bst+[\mu^{-1}]_\alpha)} 
} \nonumber\\
&&\qquad\qquad\qquad
= \frac{\lambda\mu}{\lambda-\mu}\rd_{01}\log
  \frac{\tau(\bss-\bse_0+\bse_\alpha,\,\bst+[\mu^{-1}]_\alpha)}
       {\tau(\bss-\bse_0+\bse_\alpha,\,\bse+[\lambda^{-1}]_\alpha)},
\label{dF-eq3}\\
\lefteqn{
\epsilon_{\alpha\beta}
\frac{\tau(\bss-2\bse_0+\bse_\alpha+\bse_\beta,\,
      \bst+[\lambda^{-1}]_\alpha+[\mu^{-1}]_\beta)\tau(\bse,\bst)}
     {\tau(\bss-\bse_0+\bse_\alpha,\,\bst+[\lambda^{-1}]_\alpha)
      \tau(\bss-\bse_0+\bse_\beta,\,\bst+[\mu^{-1}]_\beta)} 
} \nonumber\\
&&\qquad\qquad\qquad
= \rd_{01}\log
  \frac{\tau(\bss-\bse_0+\bse_\beta,\,\bst+[\mu^{-1}]_\beta)}
       {\tau(\bss-\bse_0+\bse_\alpha,\,\bst+[\lambda^{-1}]_\alpha)}. 
\label{dF-eq4}
\eeq
We now convert these equations to auxiliary linear 
equations for the scalar-valued wave functions 
\beq
\begin{aligned}
\Psi_0(\bss,\bst,z) 
=& \frac{\tau(\bss,\,\bst-[z^{-1}]_0)}{\tau(\bss,\bst)}
   z^{s_0}e^{\xi(\bst_0,z)}, \\
\Psi_\beta(\bss,\bst,z) 
=& \tilde{\epsilon}_\beta(\bss) 
   \frac{\tau(\bss+\bse_0-\bse_\beta,\,\bst-[z^{-1}]_\beta)}
   {\tau(\bss,\bst)}
   z^{s_\beta}e^{\xi(\bst_\beta,z)}.
\end{aligned}
\label{Psi-def}
\eeq
To simplify notations, these wave functions are 
referred to as $\Psi_0(z)$ and $\Psi_\beta(z)$.  
We have the following analogue of 
(\ref{KP-gen-lineq}) for these wave functions.  

\begin{theorem}
(\ref{dF-eq1})--(\ref{dF-eq4}) can be converted 
to the linear equations 
\beq
  \lambda e^{-D(\lambda)}\Psi(\mu) 
  &=& - (\rd_{01} - \rd_{01}\log\Psi_0(\lambda))\Psi(\mu), 
\label{gen-lineq1} \\
  e^{-\rd_{\alpha 0}-D_\alpha(\lambda)}\Psi(\mu) 
  &=& (\rd_{01} - \rd_{01}\log\Psi_\alpha(\lambda))\Psi(\mu) 
      \quad (\alpha = 1,\ldots,N) 
\label{gen-lineq2}
\eeq
for $\Psi(\mu) = \Psi_0(\mu),
\Psi_1(\mu),\ldots,\Psi_N(\mu)$. 
\end{theorem}

\proof
Let us first consider (\ref{dF-eq1}).  
This equation has the same structure 
as the differential Fay identity of 
the KP hierarchy.  Therefore we can 
repeat the foregoing calculations 
for the KP hierarchy to derive the equation 
\beqnn
  \lambda e^{-D_0(\lambda)}\Psi_0(\mu) 
  = - (\rd_{01} - \rd_{01}\Psi_0(\lambda))\Psi_0(\mu), 
\eeqnn
which is nothing but (\ref{gen-lineq1}) 
for $\Psi(\mu) = \Psi_0(\mu)$.  

As regards (\ref{dF-eq2}), we shift 
the variables as 
\beqnn
  \bss \to \bss + \bse_0 - \bse_\alpha, \quad 
  \bst \to \bst - [\lambda^{-1}] - [\mu^{-1}] 
\eeqnn
and consider the equation 
\beqnn
\begin{aligned}
 &\lambda\frac{\tau(\bss+\bse_0-\bse_\alpha,\,
         \bst-[\lambda^{-1}]_0-[\mu^{-1}]_\alpha)\tau(\bss,\bst)} 
   {\tau(\bss,\,\bst-[\lambda^{-1}]_0)
    \tau(\bss+\bse_0-\bse_\alpha,\,\bst-[\mu^{-1}]_\alpha)}\\
 &= \lambda 
    + \rd_{01}\log
      \frac{\tau(\bss,\,\bst-[\lambda^{-1}]_0)} 
      {\tau(\bss+\bse_0-\bse_\alpha,\,\bst-[\mu^{-1}]_\alpha)} 
\end{aligned}
\eeqnn
thus obtained.  We can rewrite this equation 
in two different ways as 
\beqnn
\begin{aligned}
 &\lambda e^{-\rd_{\alpha 0}-D_\alpha(\mu)}
  \frac{\tau(\bss,\,\bst-[\lambda^{-1}]_0)}{\tau(\bss,\bst)}\\
 &= \frac{\tau(\bss,\,\bst-[\lambda^{-1}]_\alpha)}{\tau(\bss,\bst)} 
    \Bigl(\lambda + \rd_{01}\log
      \frac{\tau(\bss+\bse_0-\bse_\alpha,\,\bst-[\lambda^{-1}]_0)}
           {\tau(\bss,\,\bst-[\mu^{-1}]_\alpha)} \Bigr) 
\end{aligned}
\eeqnn
and 
\beqnn
\begin{aligned}
 &\lambda e^{-D_0(\lambda)}
  \frac{\tau(\bss+\bse_0-\bse_\alpha,\,\bst-[\mu^{-1}]_\alpha)}
       {\tau(\bss,\bst)} \\
 &= \frac{\tau(\bss+\bse_0-\bse_\alpha,\,\bst-[\mu^{-1}]_\alpha)}
         {\tau(\bss,\bst)} 
    \Bigl(\lambda + \rd_{01}\log 
      \frac{\tau(\bss+\bse_0-\bse_\alpha,\,\bst-[\lambda^{-1}]_0)}
           {\tau(\bss,\,\bst-[\mu^{-1}]_\alpha)} \Bigr). 
\end{aligned}
\eeqnn
Upon interchanging $\lambda$ and $\mu$, 
the first equation reduces to 
\beqnn
  e^{-\rd_{\alpha 0}-D_\alpha(\lambda)}\Psi_0(\mu) 
  = (\rd_{01} - \rd_{01}\log\Psi_\alpha(\lambda))\Psi_0(\mu),
\eeqnn
which coincides with (\ref{gen-lineq2}) 
for the case of $\Psi(\mu) = \Psi_0(\mu)$.  
On the other hand, the second equation becomes 
\beqnn
  \lambda e^{-D_0(\lambda)}\Psi_\alpha(\mu) 
  = - (\rd_{01} - \rd_{01}\log\Psi_0(\lambda))\Psi_\alpha(\mu), 
\eeqnn
yielding (\ref{gen-lineq1}) for the case of 
$\Psi(\mu) = \Psi_\alpha(\mu)$.  

We now turn to (\ref{dF-eq3}).  
By shifting the variables as 
\beqnn
  \bss \to \bss + 2\bse_0 - 2\bse_\alpha, \quad 
  \bst \to \bst - [\lambda^{-1}]_\alpha - [\mu^{-1}]_\alpha, 
\eeqnn
we obtain the equation 
\beqnn
\begin{aligned}
 & \frac{\tau(\bss+2\bse_0-2\bse_\alpha,\, 
         \bst-[\lambda^{-1}]_\alpha-[\mu^{-1}]_\alpha) \tau(\bss,\bst)}
        {\tau(\bss+\bse_0-\bse_\alpha,\,\bst-[\lambda^{-1}]_\alpha)
        \tau(\bss+\bse_0-\bse_\alpha,\,\bst-[\mu^{-1}]_\alpha)}\\
 &= \frac{\lambda\mu}{\lambda-\mu}\rd_{01}\log 
    \frac{\tau(\bss+\bse_0-\bse_\alpha,\,\bst-[\lambda^{-1}]_\alpha)}
         {\tau(\bss+\bse_0-\bse_\alpha,\,\bst-[\mu^{-1}]_\alpha)},  
\end{aligned}
\eeqnn
equivalently, 
\beqnn
\begin{aligned}
 &e^{-\rd_{\alpha 0}-D_\alpha(\lambda)}
  \frac{\tau(\bss+\bse_0-\bse_\alpha,\,\bst-[\mu^{-1}]_\alpha)}
       {\tau(\bss,\bst)} \\
 &= \frac{\tau(\bss+\bse_0-\bse_\alpha,\,\bst-[\mu^{-1}]_\alpha)}
         {\tau(\bss,\bst)}
    \frac{\lambda\mu}{\lambda-\mu}\rd_{01}\log
    \frac{\tau(\bss+\bse_0-\bse_\alpha,\,\bst-[\lambda^{-1}]_\alpha)}
         {\tau(\bss+\bse_0-\bse_\alpha,\,\bst-[\mu^{-1}]_\alpha)}. 
\end{aligned}
\eeqnn
We now use the difference relation 
\beqnn
  \tilde{\epsilon}_\alpha(\bss+\bse_0-\bse_\alpha) 
  = - \tilde{\epsilon}_\alpha(\bss) 
\eeqnn
to convert the last equation to the linear equation 
\beqnn
  e^{-\rd_{\alpha 0}-D_\alpha(\lambda)}\Psi_\alpha(\mu) 
  = (\rd_{01} - \rd_{01}\log\Psi_\alpha(\lambda))\Psi_\alpha(\mu) 
\eeqnn
for $\Psi_\alpha(\mu)$.  This is exactly 
(\ref{gen-lineq2}) for the case of 
$\Psi(\mu) = \Psi_\alpha(\mu)$.  

Lastly, we consider (\ref{dF-eq4}).  
Upon shifting variables as 
\beqnn
  \bss \to \bss + 2\bse_0 - \bse_\alpha - \bse_\beta, \quad 
  \bst \to \bst - [\lambda^{-1}]_\alpha - [\mu^{-1}]_\beta, 
\eeqnn
we have the equation 
\beqnn
\begin{aligned}
 &\epsilon_{\alpha\beta}
  \frac{\tau(\bss+2\bse_0-\bse_\alpha-\bse_\beta,\,
        \bst-[\lambda^{-1}]_\alpha-[\mu^{-1}]_\beta)\tau(\bss,\bst)}
       {\tau(\bss+\bse_0-\bse_\alpha,\,\bst-[\lambda^{-1}]_\alpha)
        \tau(\bss+\bse_0-\bse_\beta,\,\bst-[\mu^{-1}]_\beta)}\\
 &= \rd_{01}\log
    \frac{\tau(\bss+\bse_0-\bse_\alpha,\,\bst-[\lambda^{-1}]_\alpha)}
         {\tau(\bss+\bse_0-\bse_\beta,\,\bst-[\mu^{-1}]_\beta)}, 
\end{aligned}
\eeqnn
equivalently, 
\beqnn
\begin{aligned}
 &\epsilon_{\alpha\beta}
  e^{-\rd_{\alpha 0}-D_\alpha(\lambda)} 
  \frac{\tau(\bss+\bse_0-\bse_\beta,\,\bst-[\mu^{-1}]_\beta)}
       {\tau(\bss,\bst)} \\
 &= \frac{\tau(\bss+\bse_0-\bse_\beta,\,\bst-[\mu^{-1}]_\beta)}
         {\tau(\bss,\bst)} 
    \rd_{01}\log
    \frac{\tau(\bss+\bse_0-\bse_\alpha,\,\bst-[\lambda^{-1}]_\alpha)}
         {\tau(\bss+\bse_0-\bse_\beta,\,\bst-[\mu^{-1}]_\beta)}. 
\end{aligned}
\eeqnn
Noting the difference relation 
\beqnn
  \tilde{\epsilon}_\beta(\bss+\bse_0-\bse_\alpha) 
  = - \epsilon_{\alpha\beta}\tilde{\epsilon}_\beta(\bss), 
\eeqnn
we can convert the last equation to the linear equation 
\beqnn
  e^{-\rd_{\alpha 0}-D_\alpha(\lambda)}\Psi_\beta(\mu) 
  = (\rd_{01} - \rd_{01}\log\Psi_\alpha(\lambda))\Psi_\beta(\mu)
\eeqnn
for $\Psi_\beta(\mu)$.  This gives (\ref{gen-lineq2}) 
for $\Psi(\mu) = \Psi_\beta(\mu)$, $\beta \not= \alpha$.  
Thus all equations of (\ref{gen-lineq1}) and 
(\ref{gen-lineq2}) have been derived from 
the multi-component differential Fay identities. 
\qed

\subsection{Evolutionary form of auxiliary linear equations}

We can convert the generating functional auxiliary 
linear equations (\ref{gen-lineq1})--(\ref{gen-lineq2}) 
to an ``evolutionary'' form.  As it turns out below, 
those auxiliary linear equations turn out to be 
a mixture of those of one-component KP hierarchies 
and two-dimensional Toda field equations.  
The KP hierarchies live in each sector of 
the $N+1$ sets of time variables $\bst_0,\bst_1,\ldots,\bst_N$. 
The Toda field equations (and, actually, the Toda hierarchies) 
connect the $\bst_0$-sector with the other $N$ sectors 
pairwise.  We can thus eventually recover all building blocks 
of the scalar Lax formalism of the multi-component 
KP hierarchy \cite{Takasaki-SISSA05}.

\subsubsection{KP hierarchy in $\bst_0$-sector}

Let us note that the first set (\ref{gen-lineq1}) 
of these equations have the same structure as 
their counterpart (\ref{KP-gen-lineq}) 
in the KP hierarchy.  Therefore they can be 
converted to linear equations of the form 
\beq
  \rd_{0n}\Psi(z) = B_{0n}(\rd_{01})\Psi(z) 
\label{KP-lineq0}
\eeq
for $\Psi(z) = \Psi_0(z),\Psi_1(z),\ldots,\Psi_N(z)$. 
$B_{0n}(\rd_{01})$ is a differential operator 
of the form 
\beqnn
  B_{0n}(\rd_{01}) = \rd_{01}^n + O(\rd_{01}^{n-2}). 
\eeqnn
Among the $N+1$ wave functions, $\Psi_0(z)$ 
plays the role of the wave function 
in the usual KP hierarchy.  If we define 
the dressing operator 
\beqnn
 W_0(\rd_{01}) 
  = 1 + \sum_{j=1}^\infty w_{0j}\rd_{01}^{-j} 
\eeqnn
such that 
\beqnn
  \Psi_0(z) 
  = W_0(\rd_{01})z^{s_0}e^{\xi(\bst_0,z)} 
  = \Bigl(1 + \sum_{j=1}^\infty w_{0j}z^{-j}\Bigr) 
    z^{s_0}e^{\xi(\bst_0,z)}, 
\eeqnn
the auxiliary linear equations for $\Psi_0(z)$ 
can be converted to the so called Sato equations 
\beq
  \frac{\rd W_0(\rd_{01})}{\rd t_{0n}} 
  = B_{0n}(\rd_{01})W_0(\rd_{01}) 
    - W_0(\rd_{01})\rd_{01}^n.  
\eeq
As usual, this implies that $B_{0n}(\rd_{01})$ 
is given by the KP-like formula 
\beq
  B_{0n}(\rd_{01}) 
  = \bigl(L_0(\rd_{01})^n\bigr)_{\ge 0}, 
\eeq
where $(\quad)_{\ge 0}$ denotes the projection 
to nonnegative powers of $\rd_{01}$, and 
$L_0(\rd_{01})$ is the pseudo-differential 
Lax operator 
\beqnn
  L_0(\rd_{01}) 
  = W_0(\rd_{01})\cdot\rd_{01}\cdot W_0(\rd_{01})^{-1}. 
\eeqnn

\subsubsection{Toda field equation in $(\bst_0,\bst_\alpha)$-sector}

We now turn to (\ref{gen-lineq2}).  As we show below, 
the auxiliary linear equations of Toda fields emerge 
in the lowest and next-to-lowest orders of expansion 
in powers of $\lambda$.  

The Toda fields $\phi_\beta = \phi_\beta(\bss,\bst)$ 
are defined in an exponentiated form as 
\beq
  e^{\phi_\beta}
  = \tilde{\epsilon}_\beta(\bss) 
    \frac{\tau(\bss+\bse_0-\bse_\beta,\,\bst)}{\tau(\bss,\bst)}, 
\eeq
which is actually the leading coefficients 
of the amplitude part of $\Psi_\beta(z)$, i.e., 
\beqnn
  \Psi_\beta(z) 
  = \Bigl(e^{\phi_\beta} 
      + \sum_{j=1}^\infty w_{\beta j}z^{-j} \Bigr)
    z^{s_\beta}e^{\xi(\bst_\beta,z)}. 
\eeqnn

Let us examine the lowest orders of expansion of 
(\ref{gen-lineq2}) in powers of $\lambda$.  
The $\lambda^0$-terms yield a linear equation 
of the form 
\beqnn
  e^{-\rd_{\alpha 0}}\Psi_\beta(z) 
  = (\rd_{01} - \rd_{01}\phi_\alpha)\Psi_\beta(z). 
\eeqnn
($\mu$ has been replaced by $z$.) 
Introducing a new field $q_\alpha$ as 
\beq
  q_\alpha = \rd_{01}\phi_\alpha, 
\label{q-KP=der-phi}
\eeq
we can rewrite this equation as 
\beq
  \rd_{01}\Psi_\beta(z) 
  = (e^{-\rd_{\alpha 0}} + q_\alpha)\Psi_\beta(z).  
\label{Toda-lineq0}
\eeq
Let us note that the associated Hamilton-Jacobi equation 
\beqnn
  \rd_{01}S_\beta(z) 
  = e^{-\rd_{\alpha 0}S_\beta(z)} + q_\alpha, 
\eeqnn
in quasi-classical approximation is nothing but 
(\ref{HJ-eq-n=0}).  The $\lambda^1$-terms of 
(\ref{gen-lineq2}) give the somewhat ugly equation 
\beqnn
  -e^{-\rd_{\alpha 0}}\rd_{\alpha 1}\Psi_\beta(z) 
  = \rd_{01}\rd_{\alpha 1}
    \log\tau(\bss+\bse_0-\bse_\alpha,\,\bst) 
    \cdot \Psi_\beta(z). 
\eeqnn
We can rewrite it as 
\beq
  \rd_{\alpha 1}\Psi_\beta(z) 
  = r_\alpha e^{\rd_{\alpha 0}}\Psi_\beta(z), 
\label{Toda-lineqa}
\eeq
where 
\beq
  r_\alpha = - \rd_{01}\rd_{\alpha 1}\log\tau(\bss,\bst). 
\label{r-KP=der-phi}
\eeq
The associated Hamilton-Jacobi equation reads 
\beqnn
  \rd_{\alpha 1}S_\beta(z) 
  = r_\alpha e^{\rd_{\alpha 0}S_\beta(z)}. 
\eeqnn
Using the foregoing Hamilton-Jacobi equation 
for the $t_{01}$ \-flow, we can rewrite it as 
\beqnn
  \rd_{\alpha 1}S_\beta(z) 
  = \frac{r_\alpha}{\rd_{01}S_\beta(z) - q_\alpha}. 
\eeqnn
This coincides with (\ref{HJ-eq}) for $n = 1$.  

As one can thus see from quasi-classical approximation,  
the auxiliary fields $q_\alpha$ and $r_\alpha$ 
correspond to those of the universal Whitham hierarchy 
defined by (\ref{qr=der-phi}).  Moreover, substituting 
\beqnn
  \Psi_\alpha(z) 
  = (e^{\phi_\alpha} + O(z^{-1}))
    z^{s_\alpha}e^{\xi(\bst_\alpha,z)} 
\eeqnn
in (\ref{Toda-lineqa}) for the case of 
$\beta = \alpha$ and extracting the coefficient 
of $z^{s_\alpha}e^{\xi(\bst_\alpha,z)}$, 
we obtain another expression 
\beq
  r_\alpha 
  = \exp\bigl(\phi_\alpha(\bss,\bst) 
            - \phi_\alpha(\bss-\bse_0+\bse_\alpha,\bst)\bigr)
\label{r-KP=exp-phi}
\eeq
of $r_\alpha$.  This expression corresponds to 
(\ref{r=exp-phi}).  

(\ref{Toda-lineq0}) and (\ref{Toda-lineqa}) 
may be thought of as auxiliary linear equations 
of the two-dimensional Toda fields 
with continuous variables $t_{01},t_{\alpha 1}$ 
and discrete variable $s_\alpha$.   
The zero-curvature equation 
\beqnn
  \bigl[\rd_{01} - e^{-\rd_{\alpha 0}} - q_\alpha,\,
  \rd_{\alpha 1} - r_\alpha e^{\rd_{\alpha 0}}\bigr] 
  = 0 
\eeqnn
comprises the two equations 
\beq
\begin{aligned}
  \rd_{01}r_\alpha(\bss,\bst) 
  - (q_\alpha(\bss,\bst) 
     - q_\alpha(\bss-\bse_0+\bse_\alpha,\,\bst))
    r_\alpha(\bss,\bst) &= 0, \\
  \rd_{\alpha 1}q_\alpha(\bss,\bst) 
  + r_\alpha(\bss+\bse_0-\bse_\alpha,\,\bst) 
  - r_\alpha(\bss,\bst) &= 0.  
\end{aligned}
\label{zc-eq}
\eeq
The first equation of (\ref{zc-eq}) is automatically 
satisfied under the definition of $q_\alpha$ and 
$r_\alpha$ by (\ref{q-KP=der-phi}) and 
(\ref{r-KP=exp-phi});  the second equation 
reduces to the Toda field equation 
\beq
\begin{aligned}
 \rd_{01}\rd_{\alpha 1}\phi_\alpha(\bss,\bst) 
 &+ \exp\bigl(\phi_\alpha(\bss+\bse_0-\bse_\alpha,\,\bst) 
      - \phi_\alpha(\bss,\bst)\bigr) \\
 &\mbox{} 
  - \exp\bigl(\phi_\alpha(\bss,\bst) 
      - \phi_\alpha(\bss-\bse_0+\bss_\alpha,\,\bst)\bigr)
  = 0. 
\end{aligned}
\eeq
This explains why a multi-dimensional dispersionless 
Toda field equation (or the Boyer-Finley equation) 
shows up in the universal Whitham hierarchy 
\cite{MMMA-0509}.

\subsubsection{Auxiliary linear equation of 
Schr\"odinger type in $(\bst_0,\bst_\alpha)$-sector} 

The $(\bst_0,\bst_\alpha)$-sector is also 
accompanied by an auxiliary linear equation 
of the form 
\beq
  ((\rd_{01} - q_\alpha)\rd_{\alpha 1} - r_\alpha)\Psi(z) = 0 
\label{0a-lineq}
\eeq
satisfied by $\Psi(z) = \Psi_0(z),\Psi_1(z),\ldots,\Psi_N(z)$.  
These are analogues of two-dimensional ``integrable 
Schr\"odinger equations'' \cite{NV-84,Krichever-05}.  

(\ref{0a-lineq}) can be derived from the Toda-like 
auxiliary linear equations (\ref{Toda-lineq0}) 
and (\ref{Toda-lineqa}) as follows.   
We first apply the operator $\rd_{01} - q_\alpha$ to 
both hand sides of (\ref{Toda-lineqa}).  
This yields an equation of the form 
\beqnn
\begin{aligned}
   (\rd_{01} - q_\alpha)\Psi_\beta(z) 
&= (\rd_{01} - q_\alpha)(r_\alpha e^{\rd_{\alpha 0}}\Psi_\beta(z)) \\
&= (\rd_{01}r_\alpha)e^{\rd_{\alpha 0}}\Psi_\beta(z) 
  + r_\alpha(\rd_{01} - q_\alpha)(e^{\rd_{\alpha 0}}\Psi_\beta(z)). 
\end{aligned}
\eeqnn
By the first equation of (\ref{zc-eq}), 
we can rewrite the first term in the last line as 
\beqnn
  (\rd_{01}r_\alpha)e^{\rd_{\alpha 0}}\Psi_\beta(z) 
  = (q_\alpha(\bss,\bst) - q_\alpha(\bss-\bse_0+\bse_\alpha,\,\bst))
    r_\alpha e^{\rd_{\alpha 0}}\Psi_\beta(z). 
\eeqnn
As regards the second term, we use (\ref{Toda-lineq0}) 
to rewrite its main part as 
\beqnn
\begin{aligned}
   (\rd_{01} - q_\alpha)(e^{\rd_{\alpha 0}}\Psi_\beta(z)) 
&= e^{\rd_{\alpha 0}}\rd_{01}\Psi_\beta(z) 
   - q_\alpha e^{\rd_{\alpha 0}}\Psi_\beta(z) \\
&= e^{\rd_{\alpha 0}}(e^{-\rd_{\alpha 0}} + q_\alpha)\Psi_\beta 
   - q_\alpha\Psi_\beta(z) \\
&= \Psi_\beta(z) 
   + (q_\alpha(\bss-\bse_0+\bse_\alpha,\,\bst) - q_\alpha(\bss,\bst))
     e^{\rd_{\alpha 0}}\Psi_\beta(z). 
\end{aligned}
\eeqnn
Thus the terms proportional to $q_\alpha(\bss,\bst) 
- q_\alpha(\bss-\bse_0+\bse_\alpha,\,\bst)$ cancel 
each other, and (\ref{0a-lineq}) follows.

\subsubsection{KP hierarchies in other sectors} 

Let us now consider (\ref{gen-lineq2}).  
Applying the shift operator $e^{\rd_{\alpha 0}}$ 
to both hand side yields 
\beqnn
  e^{-D_\alpha(\lambda)}\Psi(\mu) 
  = (\rd_{01} - e^{\rd_{\alpha 0}}\rd_{01}\log\Psi_\alpha(\lambda))
    e^{\rd_{\alpha 0}}\Psi(\mu). 
\eeqnn
On the other hand, (\ref{Toda-lineqa}) implies that 
\beqnn
  e^{\rd_{\alpha 0}}\Psi(\mu) 
  = r_{\alpha}^{-1}\rd_{\alpha 1}\Psi(\mu). 
\eeqnn
Therefore we can eliminate the difference term 
$e^{\rd_{\alpha 0}}\Psi(\mu)$ and obtain 
the differential equation 
\beq
  e^{-D_\alpha(\lambda)}\Psi(\mu) 
  = (\rd_{01} - e^{\rd_{\alpha 0}}\rd_{01}\log\Psi_\alpha(\lambda))
    r_\alpha^{-1}\rd_{\alpha 1}\Psi(\mu). 
\eeq
Letting as $\lambda \to \infty$, we have 
\beqnn
  \Psi(\mu) 
  = (\rd_{01} - e^{\rd_{\alpha 0}}q_\alpha) 
    r_\alpha^{-1}\rd_{\alpha 1}\Psi(\mu). 
\eeqnn
Taking the difference of these two equations 
and doing some algebra, we eventually obtain 
an equation of the form 
\beq
  (1 - e^{-D_\alpha(\lambda)})\Psi(\mu) 
  = X_\alpha(\lambda)\Psi(\mu), 
\eeq
where $X_\alpha(\lambda)$ is the first-order 
differential operator 
\beqnn
  X_\alpha(\lambda) 
  = \Bigl(\rd_{01}\log
    \frac{\tau(\bss-\bse_0+\bse_\alpha,\,\bst-[\lambda^{-1}])}
         {\tau(\bss-\bse_0+\bse_\alpha,\,\bst)} 
    \Bigr)r_\alpha^{-1}\rd_{\alpha 1}. 
\eeqnn
We can now repeat the reasoning in the derivation 
of (\ref{KP-lineq}) from (\ref{KP-gen-lineq}).  
The outcome is a system of ``evolutionary'' 
linear equations of the form 
\beq
  \rd_{\alpha n}\Psi(z) 
  = B_{\alpha n}(\rd_{\alpha 1})\Psi(z), 
\label{KP-lineqa}
\eeq
where $B_{\alpha n}(\rd_{\alpha 1})$ is 
an $n$-th order differential operator 
without $0$-th order term, i.e., 
\beqnn
  B_{\alpha n}(\rd_{\alpha 1}) 
  = b_{\alpha n0}\rd_{\alpha 1}^n 
    + \cdots + b_{\alpha nn-1}\rd_{\alpha 1}. 
\eeqnn
The absence of $0$-th order term is 
a consequence of the structure of 
$X_\alpha(z)$.   (\ref{KP-lineq}) may be 
interpreted as auxiliary linear equations 
of the one-component KP hierarchy in 
the $\bst_\alpha$-sector.

\section{Conclusion}

We have thus identified the universal 
Whitham hierarchy of genus zero with 
the dispersionless limit of the charged 
multi-component KP hierarchy.  
In the course of this rather lengthy 
consideration, we have seen a number of 
remarkable aspects of these systems 
in themselves.  

A main conclusion of this consideration 
is that the dispersionless Hirota equations 
and the differential Fay identities 
can play the role of {\em master equations} 
that characterize the (usual and dispersionless) 
$\tau$-functions of these integrable systems.  
This interpretation is complementary to 
the conventional point of view based on 
the Hamilton-Jacobi equations and 
the auxiliary linear equations.  
We have an honest impression that 
the approach from the Hirota equations 
is technically simpler and conceptually 
more essential.  This approach deserves 
to be pursued further.  

Alongside these fundamental aspects, 
we expect some applications of the results 
of this paper.  For instance, 
the recent paper \cite{BSSST-SFT} of Bonora et al. 
conjectures the relevance of a multi-component 
version of the disperesionless Toda hierarchy 
in light-cone string field theory.  
This conjecture might be explained in our framework.  
Another intriguing issue is to generalize 
the notion of associativity equations \cite{BMRWZ-01} 
to the multi-component setting.  In this respect, 
we should mention the recent work of 
Konopelchenko and Magri \cite{KM-0606}, in which 
a novel approach to the universal Whitham hierarchy 
is discussed along with some other issues 
on dispersionless integrable systems.

\subsection*{Acknowedgements}

We would like to thank A. Zabrodin 
for allowing us to contribute this paper.  
K.T. is grateful to B. Konopelchenko, 
L. Mart\'{\i}nez Alonso, M. Ma\~{n}as and A. Sorin 
for useful discussions during the SISSA conference 
``Riemann-Hilbert Problems, Integrability and 
Asymptotics'' in September, 2005.  
This research was partially supported by 
Grant-in-Aid for Scientific Research 
No. 16340040, No. 18340061 and No. 18540210 from 
the Japan Society for the Promotion of Science.

\end{document}